\documentclass[amsmath,amssymb,10pt,aps,prl,reprint,notitlepage,showpacs,superscriptaddress,longbibliography]{revtex4-1}
\usepackage{graphicx}
\usepackage{calc}
\usepackage{braket}
\usepackage{ulem}   
\usepackage{slashed}
\usepackage{wasysym}
\usepackage{siunitx}
\usepackage{dsfont}
\usepackage{amsthm,amsmath,amsfonts,amssymb,verbatim,color}
\usepackage{graphicx}
\usepackage{subfigure}
\usepackage{bm}
\usepackage{epsfig,slashed}
\usepackage[T1]{fontenc}
\usepackage{ulem}
\usepackage[colorlinks=true,citecolor=blue,linkcolor=blue,urlcolor=blue]{hyperref}
\newcommand{\blue}[1]{{\textcolor{blue}{#1}}}

\begin{document}
\title{Observation of giant surface second harmonic generation coupled to nematic orders in the van der Waals antiferromagnet FePS$_3$}

\author{Zhuoliang Ni}
\affiliation{Department of Physics and Astronomy, University of Pennsylvania, Philadelphia, Pennsylvania 19104, U.S.A}
\author{Nan Huang}
\affiliation{Department of Materials Science and Engineering, University of Tennessee, Knoxville, TN 37996, U.S.A.}

\author{Amanda V. Haglund}
\affiliation{Department of Materials Science and Engineering, University of Tennessee, Knoxville, TN 37996, U.S.A.}

\author{David G. Mandrus}
\affiliation{Department of Materials Science and Engineering, University of Tennessee, Knoxville, TN 37996, U.S.A.}
\affiliation{Materials Science and Technology Division, Oak Ridge National Laboratory, Oak Ridge, TN, 37831, U.S.A.}

\author{Liang Wu}
\email{liangwu@sas.upenn.edu}
\affiliation{Department of Physics and Astronomy, University of Pennsylvania, Philadelphia, Pennsylvania 19104, U.S.A}

\begin{abstract}

Second harmonic generation has been applied to study lattice, electronic and magnetic proprieties in atomically thin materials. However, inversion symmetry breaking is usually required for the materials to generate a large signal. In this work, we report a giant second-harmonic generation that arises below the N\'eel temperature in few-layer centrosymmetric FePS$_3$. Layer-dependent study indicates the detected signal is from the second-order nonlinearity of the surface. The magnetism-induced surface second-harmonic response is two orders of magnitude larger than those reported in other magnetic systems, with the surface nonlinear susceptibility reaching 0.08--0.13 nm$^2$/V in 2 L--5 L samples. By combing linear dichroism and second harmonic generation experiments, we further confirm the giant second-harmonic generation is coupled to nematic orders formed by the three possible Zigzag antiferromagnetic domains. Our study shows that the surface second-harmonic generation is also a sensitive tool to study antiferromagnetic states in centrosymmetric atomically thin materials.

\end{abstract}

\maketitle

Due to its sensitivity to the symmetry-breaking phases, second-harmonic generation (SHG) has been applied to study many properties in van der Waals materials, including layer numbers\cite{linanolett13,Malardprb13}, crystal orientations\cite{linanolett13,kumarprb13,Malardprb13}, ferroelectric orders\cite{liunatcomm16,xiaoprl18,xue2dmater18}, charge-density waves\cite{Ficheraprb20,luoprl21}, lattice structure\cite{zhaolight16,wangnat17}, magnetic orders\cite{sunnat19,chuprl20,ninatnano21}, and multiferroic orders\cite{songarxiv21,junanolett21}. The commonly known SHG response is from a non-zero electric-dipole term from the bulk states, which requires the inversion symmetry breaking of the materials. For centrosymmetric materials,  the bulk electric-dipole term is prohibited. However, the inversion symmetry is naturally broken on the sample surface, which enables SHG responses from the surface layers\cite{bloembergenpr1962,guyotprb1986}. 

The surface SHG results from the breaking of inversion symmetry at an interface. It was detected in various metal\cite{bloembergenprl1966,leeprl1967,reifprl1991,reifprl1993}, and molecular systems\cite{chenprl1981,heinzprl82}, but has seldomly been reported in ultra-thin van der Waals materials. The bulk electric-dipole contribution is nearly proportional to the sample thickness when the sample thickness is smaller than the coherence length of the SHG light, but the surface term remains nearly independent of the thickness. In principle, a surface SHG signal could be comparable to the bulk SHG signal in ultra-thin noncentrosymmetric materials. As a result, surface SHG provides a possible way to detect emergent electronic and magnetic properties without the requirement of inversion symmetry breaking.

\begin{figure}
\centering
\includegraphics[width=0.5\textwidth]{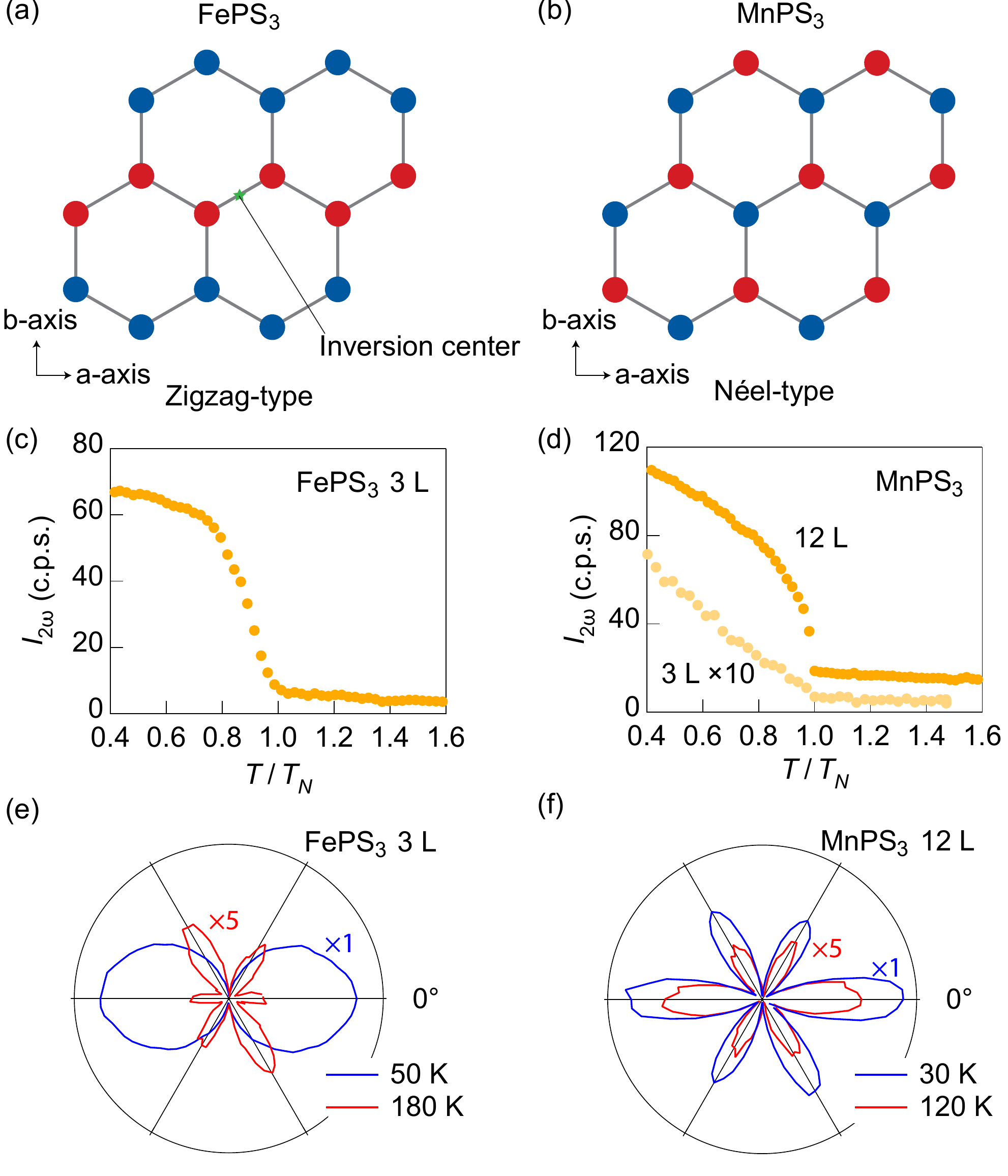}
\caption{ (a-b) Spin structure of FePS$_3$ (Zigzag-type) and MnPS$_3$ (N\'eel-type) in the monolayer. Red and blue circles represent opposite spin directions on the Fe/Mn atoms. (c-d) Temperature dependence of SHG intensity of FePS$_3$ and MnPS$_3$ samples. (e-f) Polarization-resolved SHG intensity of FePS$_3$ and MnPS$_3$ samples below and above N\'eel temperature. The polarizations of incident photons and output photons are kept parallel and rotated simultaneously.}
\label{fig1}
\end{figure}

The transition metal thiophosphates MPS$_3$ (M = Mn, Fe, Ni, Co) provide a good platform to study 2D antiferromagnetism with different magnetic anisotropy\cite{joyprb1992,Lipnas13,sivadasprb15,chittariprb16}.  Here we focus on two compounds FePS$_3$ and MnPS$_3$, with the same lattice structure. Due to the different magnetic anisotropy, FePS$_3$ has Zigzag-type antiferromagnetism where inversion symmetry is preserved\cite{kurosawajpsj1983,ruleprb07,lanconprb16} (Fig. \ref{fig1}(a)),  but MnPS$_3$ has N\'eel-type antiferromagnetism where the inversion symmetry is broken\cite{kurosawajpsj1983,ressoucheprb10} (Fig. \ref{fig1}(b)). According to the previous studies, in the N\'eel-type antiferromagnet MnPS$_3$, an electric-dipole SHG signal induced by the magnetism is observed down to bilayer samples\cite{chuprl20,niprl2021}. It is usually believed that, however, the second-harmonic response is insensitive to the Zigzag-type antiferromagnetic orders because of the preserved inversion symmetry\cite{chuprl20}.

In the monolayer FePS$_3$, the Fe atoms carrying spins form the honeycomb lattice with three-fold rotational symmetry. The space group without magnetism is $P\overline{3}1m$. In multi-layer FePS$_3$, the monoclinic stacking breaks the three-fold rotational symmetry and the space group without magnetism falls into $2/m$\cite{guyotprb1986}. When the FePS$_3$ is below the N\'eel temperature (118 K for bulk), a Zigzag-type antiferromagnetic order forms (Fig.\ref{fig1}(a)) according to the neutron scattering measurement\cite{kurosawajpsj1983,ruleprb07,lanconprb16}. The magnetism persists to the monolayer according to the Raman spectroscopy\cite{leenanoletter16,wang2dmat16}. Nevertheless, there has been some controversy over the Zigzag directions and domains. Some neutron scattering experiments suggest the Zigzag direction is along the a-axis\cite{kurosawajpsj1983} while others suggest the Zigzag direction can be 120$^\circ$ from the a-axis\cite{ruleprb07}. A more recent neutron scattering work reports the coexistence of the three Zigzag directions in a FePS$_3$ crystal, which is argued to be related with the crystal twinning rather
than magnetic domains\cite{lanconprb16}, and also concluded the structure twinning effect in the earlier work of Ref.\onlinecite{ruleprb07}. A most recent linear dichroism (LD) measurement on a multi-layer FePS$_3$ shows the co-existing of two different Zigzag directions on areas with different thickness\cite{zhangnanolett21,Xzhangnanolett21}.

In this paper, we show that the centrosymmetric Zigzag orders in FePS$_3$ can produce a surprisingly large surface second-harmonic generation down to the bilayer, which is coupled to the Zigzag antiferromagnetic orders. To investigate the origin of the multidomains, we perform spacial scanning of polarization-dependent SHG and LD measurements simultaneously on a 3L sample with a single structure domain, and observe three magnetic domains with Zigzag directions 120$^{\circ}$ to each other.   We further show that the surface SHG also rotated by 120$^{\circ}$ between these three domains. The surface SHG is coupled to the Zigzag direction, and, therefore,few-layer FePS$_3$ samples form the antiferromagnetic nematic orders without structure twinning.

 Fig. \ref{fig1}(d) show the temperature dependence of SHG intensity of a 3 L and a 12 L MnPS$_3$ flake. A sudden rise of SHG intensity below the N\'eel temperature of 78 K is observed, because the N\'eel-type antiferromagnetism breaks the inversion symmetry in this system\cite{chuprl20,niprl2021}. The SHG response is believed to have two parts, including the electric-dipole term and the electric-quadruple term:
\begin{equation}
\begin{aligned}
    {E_i}({2\omega})\propto\int_0^d \chi^{D}_{ijk}E_{j}(\omega)E_{k}(\omega)dz
        +\int_0^d\chi^{Q}_{ijzk}E_{j}(\omega)k_zE_{k}(\omega)dz,
\end{aligned}
\end{equation}
where $\chi^{D}_{ijk}$, $\chi^{Q}_{ijkl}$ are the susceptibility tensor of electric-dipole and electric-quadruple terms, respectively. $E_i(\omega)$ is the electric field from the fundamental light, and $d$ is the sample thickness.
Note that the SHG signal scales with the effective thickness as reported in the previous research\cite{niprl2021}. 

The SHG intensity of the 3 L FePS$_3$, to our surprise, is also coupled to the centrosymmetric Zigzag orders, as it shows a phase transition at the N\'eel temperature of 116 K (Fig. \ref{fig1} (c)). Moreover, the 3 L FePS$_3$ generates only slightly smaller SHG intensity than the 12 L MnPS$_3$ and 10 times larger than 3 L MnPS$_3$, which is quite surprising considering that the FePS$_3$ sample is centrosymmetric but MnPS$_3$ breaks inversion. Fig. \ref{fig1}(e) shows the polarization-dependent SHG intensity ($E_{\omega}|| E_{2\omega}$) of the same FePS$_3$ sample measured above and below N\'eel temperature. At 50 K, the SHG pattern has two lobes, which changes dramatically from the 6-lobe shape from the lattice at 180 K. Because the second-order susceptibility is directly coupled to the symmetry of the system, the arising large two-fold SHG signal in FePS$_3$ indicates that the SHG comes from the magnetism. Different from FePS$_3$,  the shapes of the SHG polar patterns of the 12 L N\'eel-type MnPS$_3$ remain the same above and below N\'eel temperature while the SHG intensity changes (Fig.\ref{fig1}(f)). 

\begin{figure}
\centering
\includegraphics[width=0.5\textwidth]{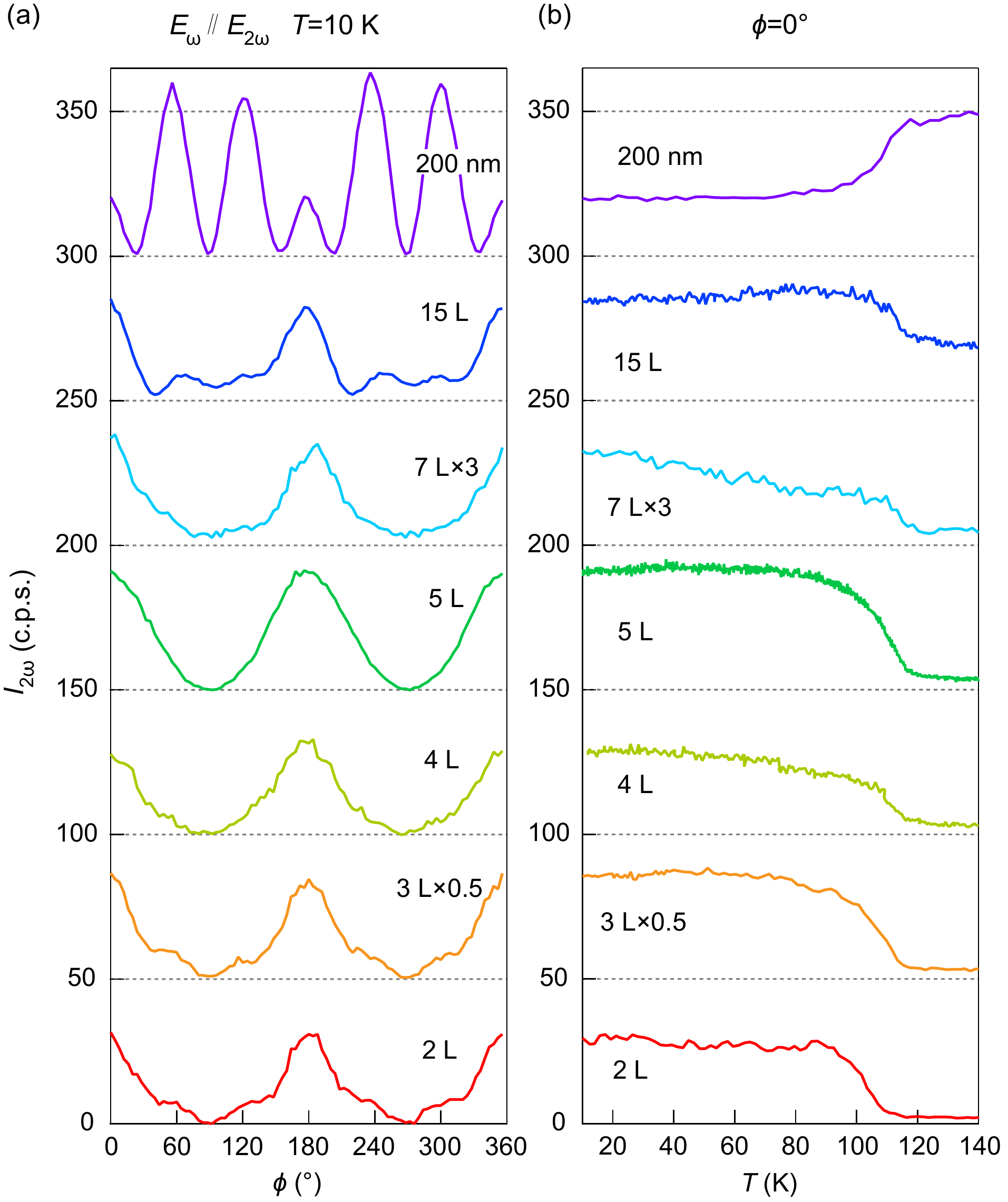}
\caption{ (a) SHG intensity as a function of polarization angle $\phi$ measured at different sample thickness. The polarizations of incident photons and output photons are kept parallel and rotated simultaneously. (b) Temperature-dependent SHG intensity measured at different sample thickness. Both polarizations of incident and output photons are kept at 0 degree.  For both (a) and (b), an offset of 50 c.p.s. is applied between different curves.}
\label{fig2}
\end{figure}

Due to the presence of the inversion symmetry in FePS$_3$, the bulk electric-dipole term is not allowed. The magnetism-coupled SHG change could come from the electric-quadruple term or/and the surface term. To better understand the origin of the large SHG response in FePS$_3$, we study the thickness dependence.  The results of polarization dependence and temperature dependence are shown in Fig. \ref{fig2}. We find that from thick flake to bilayer, the SHG intensity above N\'eel temperature drops dramatically, but the magnetism-coupled signal below $T_N$ remains in the same order of magnitude. The former is easy to understand because the SHG above N\'eel temperature is from electric-quadruple contribution, which scales quadratically on the sample thickness when the sample thickness is less than the coherence length. 
The latter indicates the magnetism-coupled SHG is insensitive to the thickness, in sharp contrast to MnPS$_3$\cite{niprl2021} and MnPSe$_3$\cite{ninatnano21} with noncentrosymmetric magnetic structures. This feature suggests that the magnetism-coupled SHG response does not originate from the bulk (such as a bulk quadruple term or a bulk magnetic dipole term). Therefore, we believe the large magnetism-coupled signal is dominated by the nonlinear response of the surface layers. The SHG response is then written as
\begin{equation}
\begin{aligned}
        E_{i}(2\omega)= \int_0^d \chi^{Q}_{ijzk}E_{j}(\omega)k_zE_{k}(\omega)dz
        +\chi^{s,z}_{ijk}E_{j}(\omega)E_{k}(\omega)|_{z=0},
\end{aligned}
\end{equation}
where $\chi^{Q}_{ijkl}$, $\chi^{s,z}_{ijk}$ are the susceptibility tensor of bulk electric-quadruple and surface terms, respectively. $E_i(\omega)$ is the electric field from the fundamental light.
The magnetism-induced surface SHG is surprisingly large. The ratio of the SHG intensity of a bilayer FePS$_3$ and a bilayer MnPS$_3$ is around 30:1 at 5 K, even though the former is centrosymmetric but the latter breaks the inversion symmetry. By using a GaAs crystal as a reference (see Supplementary materials (SM) \blue{section S2}), we show the surface nonlinear susceptibility $|\chi^{s}_{ijk}|$ reaches 0.08--0.13 nm$^2$/V (1.9--3.1$\times$10$^{-14}$ esu) in 2 L--5 L samples, which is close to the response from a noncentrosymmetric MoS$_2$ monolayer\cite{linanolett13}, whose inversion symmetry is broken by the lattice. Note that the surface SHG from FePS$_3$ is from the magnetism. When compared to other magnetism-induced surface SHG on previous reported thin-film materials\cite{reifprl1991,panprb1989,reifprl1993}, the response from FePS$_3$ is at least two orders of magnitude larger in terms of the second-order susceptibility $|\chi^{s}_{ijk}|$. Except from the large surface nonlinear optical response, it is interesting to note that FePS$_3$ also has a giant antiferromagnetism-coupled linear optical response\cite{zhangnanolett21,zhangarxiv2022}.

\begin{figure*}
\centering
\includegraphics[width=\textwidth]{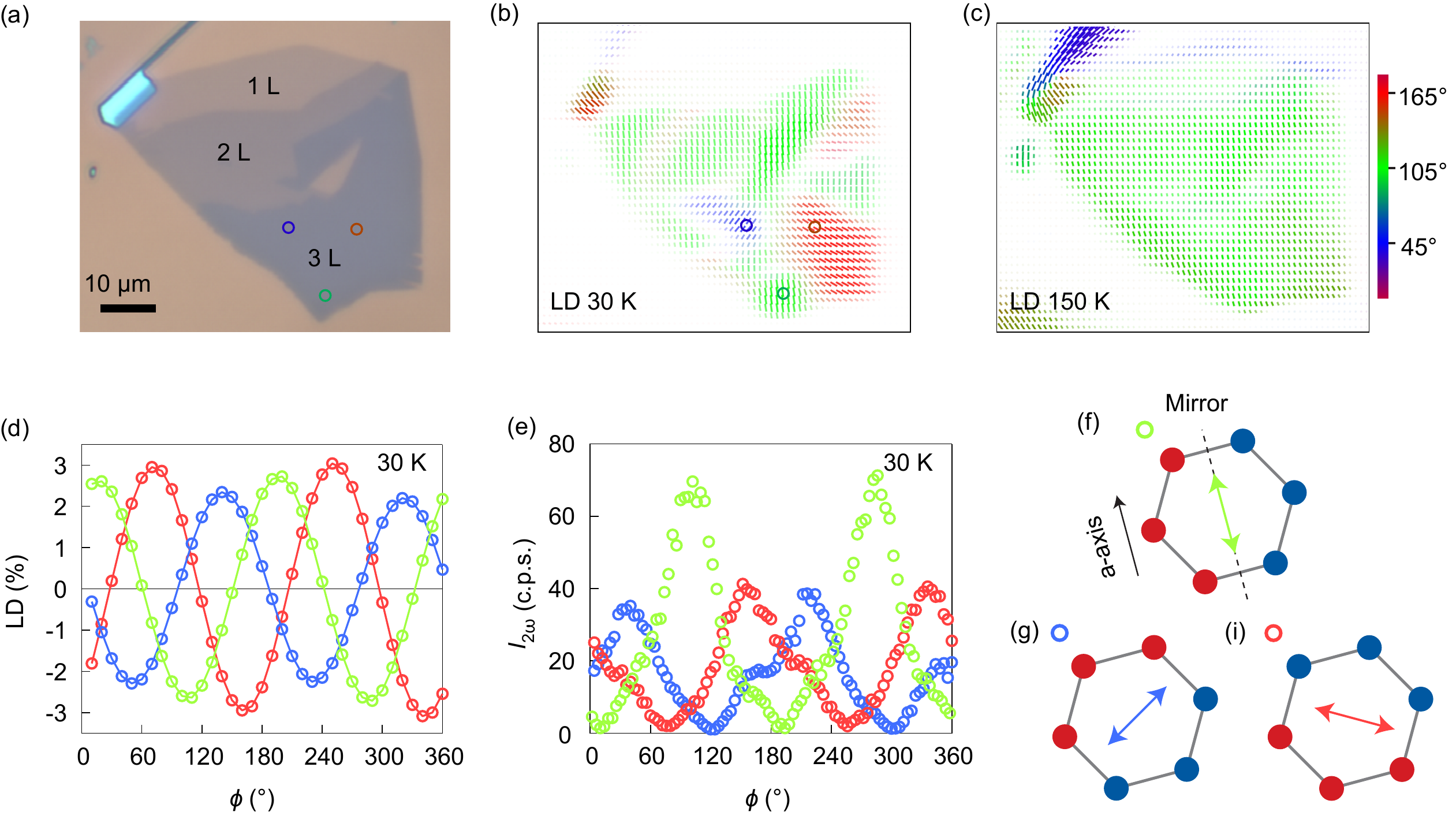}
\caption{ (a) Optical image of a multi-layer FePS$_3$ flake, including 1 L to 3 L. (b-c) LD mapping of the same area as (a) at 30 K and 150 K. The direction of small segments on each point represents the polarization direction when the LD is at the valley. The length of the segments represents the magnitude of the LD. Three colors are used to indicate three nematic domains. (d-e) Polarization-dependent LD and SHG intensity of the three points chosen from three domains marked by open circles on (b). (f-h) Schematic diagrams showing three possible nematic domains formed by Zigzag antiferromagnetic orders. Red and blue circles indicate opposite spin directions on Fe atoms. Only the (f) configuration preserves the mirror symmetry of the whole system.}
\label{fig3}
\end{figure*}

In a monolayer FePS$_3$, the Fe atoms with out-of-plane spins form the honeycomb lattice with three-fold rotational symmetry. In multi-layer FePS$_3$, the monoclinic stacking breaks the three-fold rotational symmetry very weakly. As the polarization-dependent SHG polar plots show a six-fold pattern (see SM section \blue{S1} for the high-temperature data in a bulk sample). Also, the negligible polarization-dependent LD at high temperature\cite{zhangnanolett21} suggests the system still has an approximately three-fold rotational symmetry in its electronic structure without magnetism, which gives rise to different Zigzag-order states with very close energy. The Zigzag order, on the other hand, strongly picks up a direction that breaks the approximate three-fold rotational symmetry, resulting in anisotropic linear and nonlinear optical properties that are 120$^\circ$ to each other (See Fig.\ref{fig3} and discussions below).

To address the origin of the multi-domains, we perform the LD and SHG measurements on a large 3 L FePS$_3$ sample, as thick samples often have stacking faults. See SM section \blue{S3}. Fig.\ref{fig3}(a) is the optical image of the sample, and Fig.\ref{fig3}(b) is the LD mapping at 30 K in the same region. The direction and the length of each segment at each point represent the valley location in the angle dependence and the magnitude of the LD, respectively. Three colors are used to emphasize three Zigzag directions with an interval of 120 degrees.   Three points in the green, blue and red regions of the 3L sample in Fig. \ref{fig3}(d) are chosen to show the polarization-dependent LD, which indicates the coexistence of three Zigzag directions. Note that a homogeneous LD direction distribution is observed at 150 K (Fig. \ref{fig3}(c)), excluding the influence of the stacking fault. The polarization-dependent SHG measurement at the same three points (Fig.\ref{fig3}(e)) also shows the SHG polar patterns are also rotated by 120$^\circ$, and follow the Zigzag directions. We also combine the LD and SHG measurements to show that the valley position of the LD and the peak position of the SHG indicate the Zigzag direction (see SM section \blue{S1}. It is interesting to note the intensity of the green curve is larger than the other two, and the green curve has a mirror symmetry but the red and blue ones do not. According to the LD measurement at 150 K, the Zigzag direction of the green region is along the a-axis, which preserves the mirror symmetry of the system (Fig.\ref{fig3}(f)) while the Zigzag orders of the blue (Fig.\ref{fig3}(g)) and red (Fig.\ref{fig3}(h)) region break the mirror symmetry because the Zigzag direction is different from the layer-stacking direction.

From the experimental data, it is not clear what causes different Zigzag order distribution in FePS$_3$ samples. Note the LD mapping does not change after a thermal cycle (see SM section \blue{S4}). Based on the similar behavior of other materials, we believe strain is likely responsible for the Zigzag order distribution as the three-fold rotational symmetry is only very weakly broken above $T_N$. A strain-tuning measurement as the previous study\cite{ninatnano21} by a polymer substrate is not possible due to the large and randomly distributed LD signal from the uneven polymer substrate. Similarly, the polymer substrate generates a strong photo-luminescence signal that overwhelms the SHG signal of the samples. Further experiments, for instance, adding strain directly on the wafer, are called to understand the nematic orders of the materials.

To summarize, we discover a giant surface SHG signal on FePS$_3$ samples coupled to the centrosymmetric Zigzag-type antiferromagnetic orders. The surface nonlinear susceptibility $|\chi^{s}_{ijk}|$ reaches 0.08--0.13 nm$^2$/V (1.9--3.1$\times$10$^{-14}$ esu) in 2 L--5 L samples. The direction of the Zigzag order is resolved by polarization-dependent SHG and LD measurement. Nematic states with three Zigzag directions co-exist in a single crystal, which, therefore, points to a new mechanism other than the structure twinning.
The observed giant surface SHG signal coupled to the centrosymmetric antiferromagnetic orders suggests that SHG is a good probe for emergent states in atomically-thin materials even without inversion symmetry breaking.
\\

\section{Methods}
\subsection{Sample preparation}
{To perform optical measurements in the cryostat, the bulk FePS$_3$ is mounted on a copper by a thin layer of GE varnish. The few-layer samples are directly exfoliated on the SiO$_2$/Si wafer in air and then transferred into the vacuum chamber of a cryostat. In our experiment we do not see noticeable degradation of the few-layer FePS$_3$ in air. However, we still control the exposure time to be less than 20 minutes.}

\subsection{Second harmonic generation measurement}
{In the SHG experiment, we use an 800 nm Ti-sapphire laser (80 MHz, $\sim$ 50 fs) as the fundamental light. The beam is focused by a 50$\times$ objective to the beam spot size of around 2 $\mu$m. A typical laser power of 400 $\mu$W is applied. No noticeable sample degradation is observed under the laser exposure and thermal cycles. The same objective is used to collect the reflected 400 nm beam, which is then detected by a photon counter. A half-wave plate and a linear polarizer are used to control the polarization of the incident and outgoing beams, respectively. The coherence length of the reflected SHG can be calculated by $\frac{\lambda}{4\pi(\tilde{n}_\omega+\tilde{n}_{2\omega})}$,  which is the effective thickness of the sample that can generate the in-phase reflected second-harmonic photons.}

\subsection{Linear dichroism measurement}
{The polarization-dependent LD is measured by taking the reflection difference between two axes that are perpendicular to each other: $\eta(\phi)=\frac{I_R(\phi)-I_R(\phi+90^\circ)}{I_R(\phi)+I_R(\phi+90^\circ)}$ under the same 800 nm Ti-sapphire laser as the SHG measurement. A photoelastic modulator is used to modulate the polarization at 42 kHz. A half-wave plate is placed right before the objective to control the polarization of the incident light.}

\section{Supporting Information}
{Symmetry analysis of the surface SHG of FePS$_3$. Estimation of the surface nonlinear susceptibility $\chi_{ijk}^s$ of the few-layer FePS$_3$. Zigzag domains measured in a  FePS$_3$ thick flake. Discussion on the enhancement of magnetism-coupled SHG response in thin samples. More data on linear dichroism mapping of the 3 L FePS$_3$ sample. Linear and nonlinear responses of the monolayer FePS$_3$.}

\section{Acknowledgments}
We thank J. Kikkawa, S. Parra and H. Zhang for help on the experiments, and O. Tchernyshyov for helpful discussions. The project design, data collection and analysis are supported by L.W.'s startup package at the University of Pennsylvania. The development of the SHG photon counter is supported by the ARO  under the Grant W911NF1910342. The development of the scanning imaging microscope
is partially supported by the ARO under the Grant No. W911NF2110131 and W911NF2020166, and the University Researh Foundation. Z.N.is also  partial supported from National Science Foundation supported University of Pennsylvania Materials Research Science and Engineering Center (MRSEC)(DMR-1720530). This work was carried out in part at the
Singh Center for Nanotechnology, which is supported by the
NSF National Nanotechnology Coordinated Infrastructure
Program under Grant No. NNCI-1542153. D.G.M acknowledges support from the Gordon and Betty Moore Foundation's EPiQS Initiative, Grant GBMF9069. Z.N also acknowledges support from Vagelos Institute of Energy  Science  and  Technology  graduate  fellowship  at  the  University  of  Pennsylvania.


\begin{thebibliography}{38}%
\makeatletter
\providecommand \@ifxundefined [1]{%
 \@ifx{#1\undefined}
}%
\providecommand \@ifnum [1]{%
 \ifnum #1\expandafter \@firstoftwo
 \else \expandafter \@secondoftwo
 \fi
}%
\providecommand \@ifx [1]{%
 \ifx #1\expandafter \@firstoftwo
 \else \expandafter \@secondoftwo
 \fi
}%
\providecommand \natexlab [1]{#1}%
\providecommand \enquote  [1]{``#1''}%
\providecommand \bibnamefont  [1]{#1}%
\providecommand \bibfnamefont [1]{#1}%
\providecommand \citenamefont [1]{#1}%
\providecommand \href@noop [0]{\@secondoftwo}%
\providecommand \href [0]{\begingroup \@sanitize@url \@href}%
\providecommand \@href[1]{\@@startlink{#1}\@@href}%
\providecommand \@@href[1]{\endgroup#1\@@endlink}%
\providecommand \@sanitize@url [0]{\catcode `\\12\catcode `\$12\catcode
  `\&12\catcode `\#12\catcode `\^12\catcode `\_12\catcode `\%12\relax}%
\providecommand \@@startlink[1]{}%
\providecommand \@@endlink[0]{}%
\providecommand \url  [0]{\begingroup\@sanitize@url \@url }%
\providecommand \@url [1]{\endgroup\@href {#1}{\urlprefix }}%
\providecommand \urlprefix  [0]{URL }%
\providecommand \Eprint [0]{\href }%
\providecommand \doibase [0]{http://dx.doi.org/}%
\providecommand \selectlanguage [0]{\@gobble}%
\providecommand \bibinfo  [0]{\@secondoftwo}%
\providecommand \bibfield  [0]{\@secondoftwo}%
\providecommand \translation [1]{[#1]}%
\providecommand \BibitemOpen [0]{}%
\providecommand \bibitemStop [0]{}%
\providecommand \bibitemNoStop [0]{.\EOS\space}%
\providecommand \EOS [0]{\spacefactor3000\relax}%
\providecommand \BibitemShut  [1]{\csname bibitem#1\endcsname}%
\let\auto@bib@innerbib\@empty
\bibitem [{\citenamefont {Li}\ \emph {et~al.}(2013{\natexlab{a}})\citenamefont
  {Li}, \citenamefont {Rao}, \citenamefont {Mak}, \citenamefont {You},
  \citenamefont {Wang}, \citenamefont {Dean},\ and\ \citenamefont
  {Heinz}}]{linanolett13}%
  \BibitemOpen
  \bibfield  {author} {\bibinfo {author} {\bibfnamefont {Yilei}\ \bibnamefont
  {Li}}, \bibinfo {author} {\bibfnamefont {Yi}~\bibnamefont {Rao}}, \bibinfo
  {author} {\bibfnamefont {Kin~Fai}\ \bibnamefont {Mak}}, \bibinfo {author}
  {\bibfnamefont {Yumeng}\ \bibnamefont {You}}, \bibinfo {author}
  {\bibfnamefont {Shuyuan}\ \bibnamefont {Wang}}, \bibinfo {author}
  {\bibfnamefont {Cory~R}\ \bibnamefont {Dean}}, \ and\ \bibinfo {author}
  {\bibfnamefont {Tony~F}\ \bibnamefont {Heinz}},\ }\bibfield  {title}
  {\enquote {\bibinfo {title} {{Probing symmetry properties of few-layer
  MoS$_2$ and h-BN by optical second-harmonic generation}},}\ }\href
  {https://doi.org/10.1021/nl401561r} {\bibfield  {journal} {\bibinfo
  {journal} {Nano Lett.}\ }\textbf {\bibinfo {volume} {13}},\ \bibinfo {pages}
  {3329--3333} (\bibinfo {year} {2013}{\natexlab{a}})}\BibitemShut {NoStop}%
\bibitem [{\citenamefont {Malard}\ \emph {et~al.}(2013)\citenamefont {Malard},
  \citenamefont {Alencar}, \citenamefont {Barboza}, \citenamefont {Mak},\ and\
  \citenamefont {de~Paula}}]{Malardprb13}%
  \BibitemOpen
  \bibfield  {author} {\bibinfo {author} {\bibfnamefont {Leandro~M.}\
  \bibnamefont {Malard}}, \bibinfo {author} {\bibfnamefont {Thonimar~V.}\
  \bibnamefont {Alencar}}, \bibinfo {author} {\bibfnamefont {Ana Paula~M.}\
  \bibnamefont {Barboza}}, \bibinfo {author} {\bibfnamefont {Kin~Fai}\
  \bibnamefont {Mak}}, \ and\ \bibinfo {author} {\bibfnamefont {Ana~M.}\
  \bibnamefont {de~Paula}},\ }\bibfield  {title} {\enquote {\bibinfo {title}
  {{Observation of intense second harmonic generation from MoS${}_{2}$ atomic
  crystals}},}\ }\href {\doibase 10.1103/PhysRevB.87.201401} {\bibfield
  {journal} {\bibinfo  {journal} {Phys. Rev. B}\ }\textbf {\bibinfo {volume}
  {87}},\ \bibinfo {pages} {201401} (\bibinfo {year} {2013})}\BibitemShut
  {NoStop}%
\bibitem [{\citenamefont {Kumar}\ \emph {et~al.}(2013)\citenamefont {Kumar},
  \citenamefont {Najmaei}, \citenamefont {Cui}, \citenamefont {Ceballos},
  \citenamefont {Ajayan}, \citenamefont {Lou},\ and\ \citenamefont
  {Zhao}}]{kumarprb13}%
  \BibitemOpen
  \bibfield  {author} {\bibinfo {author} {\bibfnamefont {Nardeep}\ \bibnamefont
  {Kumar}}, \bibinfo {author} {\bibfnamefont {Sina}\ \bibnamefont {Najmaei}},
  \bibinfo {author} {\bibfnamefont {Qiannan}\ \bibnamefont {Cui}}, \bibinfo
  {author} {\bibfnamefont {Frank}\ \bibnamefont {Ceballos}}, \bibinfo {author}
  {\bibfnamefont {Pulickel~M.}\ \bibnamefont {Ajayan}}, \bibinfo {author}
  {\bibfnamefont {Jun}\ \bibnamefont {Lou}}, \ and\ \bibinfo {author}
  {\bibfnamefont {Hui}\ \bibnamefont {Zhao}},\ }\bibfield  {title} {\enquote
  {\bibinfo {title} {{Second harmonic microscopy of monolayer MoS${}_{2}$}},}\
  }\href {\doibase 10.1103/PhysRevB.87.161403} {\bibfield  {journal} {\bibinfo
  {journal} {Phys. Rev. B}\ }\textbf {\bibinfo {volume} {87}},\ \bibinfo
  {pages} {161403} (\bibinfo {year} {2013})}\BibitemShut {NoStop}%
\bibitem [{\citenamefont {Liu}\ \emph {et~al.}(2016)\citenamefont {Liu},
  \citenamefont {You}, \citenamefont {Seyler}, \citenamefont {Li},
  \citenamefont {Yu}, \citenamefont {Lin}, \citenamefont {Wang}, \citenamefont
  {Zhou}, \citenamefont {Wang}, \citenamefont {He} \emph
  {et~al.}}]{liunatcomm16}%
  \BibitemOpen
  \bibfield  {author} {\bibinfo {author} {\bibfnamefont {Fucai}\ \bibnamefont
  {Liu}}, \bibinfo {author} {\bibfnamefont {Lu}~\bibnamefont {You}}, \bibinfo
  {author} {\bibfnamefont {Kyle~L}\ \bibnamefont {Seyler}}, \bibinfo {author}
  {\bibfnamefont {Xiaobao}\ \bibnamefont {Li}}, \bibinfo {author}
  {\bibfnamefont {Peng}\ \bibnamefont {Yu}}, \bibinfo {author} {\bibfnamefont
  {Junhao}\ \bibnamefont {Lin}}, \bibinfo {author} {\bibfnamefont {Xuewen}\
  \bibnamefont {Wang}}, \bibinfo {author} {\bibfnamefont {Jiadong}\
  \bibnamefont {Zhou}}, \bibinfo {author} {\bibfnamefont {Hong}\ \bibnamefont
  {Wang}}, \bibinfo {author} {\bibfnamefont {Haiyong}\ \bibnamefont {He}},
  \emph {et~al.},\ }\bibfield  {title} {\enquote {\bibinfo {title}
  {{Room-temperature ferroelectricity in CuInP$_2$S$_6$ ultrathin flakes}},}\
  }\href {https://doi.org/10.1038/ncomms12357} {\bibfield  {journal} {\bibinfo
  {journal} {Nat. Communs.}\ }\textbf {\bibinfo {volume} {7}},\ \bibinfo
  {pages} {12357} (\bibinfo {year} {2016})}\BibitemShut {NoStop}%
\bibitem [{\citenamefont {Xiao}\ \emph {et~al.}(2018)\citenamefont {Xiao},
  \citenamefont {Zhu}, \citenamefont {Wang}, \citenamefont {Feng},
  \citenamefont {Hu}, \citenamefont {Dasgupta}, \citenamefont {Han},
  \citenamefont {Wang}, \citenamefont {Muller}, \citenamefont {Martin},
  \citenamefont {Hu},\ and\ \citenamefont {Zhang}}]{xiaoprl18}%
  \BibitemOpen
  \bibfield  {author} {\bibinfo {author} {\bibfnamefont {Jun}\ \bibnamefont
  {Xiao}}, \bibinfo {author} {\bibfnamefont {Hanyu}\ \bibnamefont {Zhu}},
  \bibinfo {author} {\bibfnamefont {Ying}\ \bibnamefont {Wang}}, \bibinfo
  {author} {\bibfnamefont {Wei}\ \bibnamefont {Feng}}, \bibinfo {author}
  {\bibfnamefont {Yunxia}\ \bibnamefont {Hu}}, \bibinfo {author} {\bibfnamefont
  {Arvind}\ \bibnamefont {Dasgupta}}, \bibinfo {author} {\bibfnamefont {Yimo}\
  \bibnamefont {Han}}, \bibinfo {author} {\bibfnamefont {Yuan}\ \bibnamefont
  {Wang}}, \bibinfo {author} {\bibfnamefont {David~A.}\ \bibnamefont {Muller}},
  \bibinfo {author} {\bibfnamefont {Lane~W.}\ \bibnamefont {Martin}}, \bibinfo
  {author} {\bibfnamefont {PingAn}\ \bibnamefont {Hu}}, \ and\ \bibinfo
  {author} {\bibfnamefont {Xiang}\ \bibnamefont {Zhang}},\ }\bibfield  {title}
  {\enquote {\bibinfo {title} {{Intrinsic two-dimensional ferroelectricity with
  dipole locking}},}\ }\href {\doibase 10.1103/PhysRevLett.120.227601}
  {\bibfield  {journal} {\bibinfo  {journal} {Phys. Rev. Lett.}\ }\textbf
  {\bibinfo {volume} {120}},\ \bibinfo {pages} {227601} (\bibinfo {year}
  {2018})}\BibitemShut {NoStop}%
\bibitem [{\citenamefont {Xue}\ \emph {et~al.}(2018)\citenamefont {Xue},
  \citenamefont {Hu}, \citenamefont {Lee}, \citenamefont {Lu}, \citenamefont
  {Zhang}, \citenamefont {Tang}, \citenamefont {Han}, \citenamefont {Hsu},
  \citenamefont {Tu}, \citenamefont {Chang}, \citenamefont {Lien},
  \citenamefont {He}, \citenamefont {Zhang}, \citenamefont {Li},\ and\
  \citenamefont {Zhang}}]{xue2dmater18}%
  \BibitemOpen
  \bibfield  {author} {\bibinfo {author} {\bibfnamefont {Fei}\ \bibnamefont
  {Xue}}, \bibinfo {author} {\bibfnamefont {Weijin}\ \bibnamefont {Hu}},
  \bibinfo {author} {\bibfnamefont {Ko-Chun}\ \bibnamefont {Lee}}, \bibinfo
  {author} {\bibfnamefont {Li-Syuan}\ \bibnamefont {Lu}}, \bibinfo {author}
  {\bibfnamefont {Junwei}\ \bibnamefont {Zhang}}, \bibinfo {author}
  {\bibfnamefont {Hao-Ling}\ \bibnamefont {Tang}}, \bibinfo {author}
  {\bibfnamefont {Ali}\ \bibnamefont {Han}}, \bibinfo {author} {\bibfnamefont
  {Wei-Ting}\ \bibnamefont {Hsu}}, \bibinfo {author} {\bibfnamefont {Shaobo}\
  \bibnamefont {Tu}}, \bibinfo {author} {\bibfnamefont {Wen-Hao}\ \bibnamefont
  {Chang}}, \bibinfo {author} {\bibfnamefont {Chen-Hsin}\ \bibnamefont {Lien}},
  \bibinfo {author} {\bibfnamefont {Jr-Hau}\ \bibnamefont {He}}, \bibinfo
  {author} {\bibfnamefont {Zhidong}\ \bibnamefont {Zhang}}, \bibinfo {author}
  {\bibfnamefont {Lain-Jong}\ \bibnamefont {Li}}, \ and\ \bibinfo {author}
  {\bibfnamefont {Xixiang}\ \bibnamefont {Zhang}},\ }\bibfield  {title}
  {\enquote {\bibinfo {title} {{Room-temperature ferroelectricity in
  hexagonally layered $\alpha$-In$_2$Se$_3$ nanoflakes down to the monolayer
  limit}},}\ }\href {\doibase https://doi.org/10.1002/adfm.201803738}
  {\bibfield  {journal} {\bibinfo  {journal} {Advanced Functional Materials}\
  }\textbf {\bibinfo {volume} {28}},\ \bibinfo {pages} {1803738} (\bibinfo
  {year} {2018})}\BibitemShut {NoStop}%
\bibitem [{\citenamefont {Fichera}\ \emph {et~al.}(2020)\citenamefont
  {Fichera}, \citenamefont {Kogar}, \citenamefont {Ye}, \citenamefont
  {G\"okce}, \citenamefont {Zong}, \citenamefont {Checkelsky},\ and\
  \citenamefont {Gedik}}]{Ficheraprb20}%
  \BibitemOpen
  \bibfield  {author} {\bibinfo {author} {\bibfnamefont {Bryan~T.}\
  \bibnamefont {Fichera}}, \bibinfo {author} {\bibfnamefont {Anshul}\
  \bibnamefont {Kogar}}, \bibinfo {author} {\bibfnamefont {Linda}\ \bibnamefont
  {Ye}}, \bibinfo {author} {\bibfnamefont {Bilal}\ \bibnamefont {G\"okce}},
  \bibinfo {author} {\bibfnamefont {Alfred}\ \bibnamefont {Zong}}, \bibinfo
  {author} {\bibfnamefont {Joseph~G.}\ \bibnamefont {Checkelsky}}, \ and\
  \bibinfo {author} {\bibfnamefont {Nuh}\ \bibnamefont {Gedik}},\ }\bibfield
  {title} {\enquote {\bibinfo {title} {{Second harmonic generation as a probe
  of broken mirror symmetry}},}\ }\href {\doibase 10.1103/PhysRevB.101.241106}
  {\bibfield  {journal} {\bibinfo  {journal} {Phys. Rev. B}\ }\textbf {\bibinfo
  {volume} {101}},\ \bibinfo {pages} {241106} (\bibinfo {year}
  {2020})}\BibitemShut {NoStop}%
\bibitem [{\citenamefont {Luo}\ \emph {et~al.}(2021)\citenamefont {Luo},
  \citenamefont {Obeysekera}, \citenamefont {Won}, \citenamefont {Sung},
  \citenamefont {Schnitzer}, \citenamefont {Hovden}, \citenamefont {Cheong},
  \citenamefont {Yang}, \citenamefont {Sun},\ and\ \citenamefont
  {Zhao}}]{luoprl21}%
  \BibitemOpen
  \bibfield  {author} {\bibinfo {author} {\bibfnamefont {Xiangpeng}\
  \bibnamefont {Luo}}, \bibinfo {author} {\bibfnamefont {Dimuthu}\ \bibnamefont
  {Obeysekera}}, \bibinfo {author} {\bibfnamefont {Choongjae}\ \bibnamefont
  {Won}}, \bibinfo {author} {\bibfnamefont {Suk~Hyun}\ \bibnamefont {Sung}},
  \bibinfo {author} {\bibfnamefont {Noah}\ \bibnamefont {Schnitzer}}, \bibinfo
  {author} {\bibfnamefont {Robert}\ \bibnamefont {Hovden}}, \bibinfo {author}
  {\bibfnamefont {Sang-Wook}\ \bibnamefont {Cheong}}, \bibinfo {author}
  {\bibfnamefont {Junjie}\ \bibnamefont {Yang}}, \bibinfo {author}
  {\bibfnamefont {Kai}\ \bibnamefont {Sun}}, \ and\ \bibinfo {author}
  {\bibfnamefont {Liuyan}\ \bibnamefont {Zhao}},\ }\bibfield  {title} {\enquote
  {\bibinfo {title} {Ultrafast modulations and detection of a ferro-rotational
  charge density wave using time-resolved electric quadrupole second harmonic
  generation},}\ }\href {\doibase 10.1103/PhysRevLett.127.126401} {\bibfield
  {journal} {\bibinfo  {journal} {Phys. Rev. Lett.}\ }\textbf {\bibinfo
  {volume} {127}},\ \bibinfo {pages} {126401} (\bibinfo {year}
  {2021})}\BibitemShut {NoStop}%
\bibitem [{\citenamefont {Zhao}\ \emph {et~al.}(2016)\citenamefont {Zhao},
  \citenamefont {Ye}, \citenamefont {Suzuki}, \citenamefont {Ye}, \citenamefont
  {Zhu}, \citenamefont {Xiao}, \citenamefont {Wang}, \citenamefont {Iwasa},\
  and\ \citenamefont {Zhang}}]{zhaolight16}%
  \BibitemOpen
  \bibfield  {author} {\bibinfo {author} {\bibfnamefont {Mervin}\ \bibnamefont
  {Zhao}}, \bibinfo {author} {\bibfnamefont {Ziliang}\ \bibnamefont {Ye}},
  \bibinfo {author} {\bibfnamefont {Ryuji}\ \bibnamefont {Suzuki}}, \bibinfo
  {author} {\bibfnamefont {Yu}~\bibnamefont {Ye}}, \bibinfo {author}
  {\bibfnamefont {Hanyu}\ \bibnamefont {Zhu}}, \bibinfo {author} {\bibfnamefont
  {Jun}\ \bibnamefont {Xiao}}, \bibinfo {author} {\bibfnamefont {Yuan}\
  \bibnamefont {Wang}}, \bibinfo {author} {\bibfnamefont {Yoshihiro}\
  \bibnamefont {Iwasa}}, \ and\ \bibinfo {author} {\bibfnamefont {Xiang}\
  \bibnamefont {Zhang}},\ }\bibfield  {title} {\enquote {\bibinfo {title}
  {{Atomically Phase-Matched Second-Harmonic Generation in a 2D Crystal}},}\
  }\href {\doibase https://doi.org/10.1038/lsa.2016.131} {\bibfield  {journal}
  {\bibinfo  {journal} {Light Sci. Appl.}\ }\textbf {\bibinfo {volume} {5}},\
  \bibinfo {pages} {e16131} (\bibinfo {year} {2016})}\BibitemShut {NoStop}%
\bibitem [{\citenamefont {Wang}\ \emph {et~al.}(2017)\citenamefont {Wang},
  \citenamefont {Xiao}, \citenamefont {Zhu}, \citenamefont {Li}, \citenamefont
  {Alsaid}, \citenamefont {Fong}, \citenamefont {Zhou}, \citenamefont {Wang},
  \citenamefont {Shi}, \citenamefont {Wang} \emph {et~al.}}]{wangnat17}%
  \BibitemOpen
  \bibfield  {author} {\bibinfo {author} {\bibfnamefont {Ying}\ \bibnamefont
  {Wang}}, \bibinfo {author} {\bibfnamefont {Jun}\ \bibnamefont {Xiao}},
  \bibinfo {author} {\bibfnamefont {Hanyu}\ \bibnamefont {Zhu}}, \bibinfo
  {author} {\bibfnamefont {Yao}\ \bibnamefont {Li}}, \bibinfo {author}
  {\bibfnamefont {Yousif}\ \bibnamefont {Alsaid}}, \bibinfo {author}
  {\bibfnamefont {King~Yan}\ \bibnamefont {Fong}}, \bibinfo {author}
  {\bibfnamefont {Yao}\ \bibnamefont {Zhou}}, \bibinfo {author} {\bibfnamefont
  {Siqi}\ \bibnamefont {Wang}}, \bibinfo {author} {\bibfnamefont
  {Wu}~\bibnamefont {Shi}}, \bibinfo {author} {\bibfnamefont {Yuan}\
  \bibnamefont {Wang}},  \emph {et~al.},\ }\bibfield  {title} {\enquote
  {\bibinfo {title} {{Structural phase transition in monolayer MoTe$_2$ driven
  by electrostatic doping}},}\ }\href {https://doi.org/10.1038/nature24043}
  {\bibfield  {journal} {\bibinfo  {journal} {Nature}\ }\textbf {\bibinfo
  {volume} {550}},\ \bibinfo {pages} {487--491} (\bibinfo {year}
  {2017})}\BibitemShut {NoStop}%
\bibitem [{\citenamefont {Sun}\ \emph {et~al.}(2019)\citenamefont {Sun},
  \citenamefont {Yi}, \citenamefont {Song}, \citenamefont {Clark},
  \citenamefont {Huang}, \citenamefont {Shan}, \citenamefont {Wu},
  \citenamefont {Huang}, \citenamefont {Gao}, \citenamefont {Chen},
  \citenamefont {McGuire}, \citenamefont {Cao}, \citenamefont {Xiao},
  \citenamefont {Liu}, \citenamefont {Yao}, \citenamefont {Xu},\ and\
  \citenamefont {Wu}}]{sunnat19}%
  \BibitemOpen
  \bibfield  {author} {\bibinfo {author} {\bibfnamefont {Zeyuan}\ \bibnamefont
  {Sun}}, \bibinfo {author} {\bibfnamefont {Yangfan}\ \bibnamefont {Yi}},
  \bibinfo {author} {\bibfnamefont {Tiancheng}\ \bibnamefont {Song}}, \bibinfo
  {author} {\bibfnamefont {Genevieve}\ \bibnamefont {Clark}}, \bibinfo {author}
  {\bibfnamefont {Bevin}\ \bibnamefont {Huang}}, \bibinfo {author}
  {\bibfnamefont {Yuwei}\ \bibnamefont {Shan}}, \bibinfo {author}
  {\bibfnamefont {Shuang}\ \bibnamefont {Wu}}, \bibinfo {author} {\bibfnamefont
  {Di}~\bibnamefont {Huang}}, \bibinfo {author} {\bibfnamefont {Chunlei}\
  \bibnamefont {Gao}}, \bibinfo {author} {\bibfnamefont {Zhanghai}\
  \bibnamefont {Chen}}, \bibinfo {author} {\bibfnamefont {Michael}\
  \bibnamefont {McGuire}}, \bibinfo {author} {\bibfnamefont {Ting}\
  \bibnamefont {Cao}}, \bibinfo {author} {\bibfnamefont {Di}~\bibnamefont
  {Xiao}}, \bibinfo {author} {\bibfnamefont {Wei-Tao}\ \bibnamefont {Liu}},
  \bibinfo {author} {\bibfnamefont {Wang}\ \bibnamefont {Yao}}, \bibinfo
  {author} {\bibfnamefont {Xiaodong}\ \bibnamefont {Xu}}, \ and\ \bibinfo
  {author} {\bibfnamefont {Shiwei}\ \bibnamefont {Wu}},\ }\bibfield  {title}
  {\enquote {\bibinfo {title} {{Giant nonreciprocal second-harmonic generation
  from antiferromagnetic bilayer CrI$_3$}},}\ }\href {\doibase
  10.1038/s41586-019-1445-3} {\bibfield  {journal} {\bibinfo  {journal}
  {Nature}\ }\textbf {\bibinfo {volume} {572}},\ \bibinfo {pages} {497--501}
  (\bibinfo {year} {2019})}\BibitemShut {NoStop}%
\bibitem [{\citenamefont {Chu}\ \emph {et~al.}(2020)\citenamefont {Chu},
  \citenamefont {Roh}, \citenamefont {Island}, \citenamefont {Li},
  \citenamefont {Lee}, \citenamefont {Chen}, \citenamefont {Park},
  \citenamefont {Young}, \citenamefont {Lee},\ and\ \citenamefont
  {Hsieh}}]{chuprl20}%
  \BibitemOpen
  \bibfield  {author} {\bibinfo {author} {\bibfnamefont {Hao}\ \bibnamefont
  {Chu}}, \bibinfo {author} {\bibfnamefont {Chang~Jae}\ \bibnamefont {Roh}},
  \bibinfo {author} {\bibfnamefont {Joshua~O.}\ \bibnamefont {Island}},
  \bibinfo {author} {\bibfnamefont {Chen}\ \bibnamefont {Li}}, \bibinfo
  {author} {\bibfnamefont {Sungmin}\ \bibnamefont {Lee}}, \bibinfo {author}
  {\bibfnamefont {Jingjing}\ \bibnamefont {Chen}}, \bibinfo {author}
  {\bibfnamefont {Je-Geun}\ \bibnamefont {Park}}, \bibinfo {author}
  {\bibfnamefont {Andrea~F.}\ \bibnamefont {Young}}, \bibinfo {author}
  {\bibfnamefont {Jong~Seok}\ \bibnamefont {Lee}}, \ and\ \bibinfo {author}
  {\bibfnamefont {David}\ \bibnamefont {Hsieh}},\ }\bibfield  {title} {\enquote
  {\bibinfo {title} {{Linear magnetoelectric phase in ultrathin
  ${\mathrm{MnPS}}_{3}$ probed by optical second harmonic generation}},}\
  }\href {\doibase 10.1103/PhysRevLett.124.027601} {\bibfield  {journal}
  {\bibinfo  {journal} {Phys. Rev. Lett.}\ }\textbf {\bibinfo {volume} {124}},\
  \bibinfo {pages} {027601} (\bibinfo {year} {2020})}\BibitemShut {NoStop}%
\bibitem [{\citenamefont {Ni}\ \emph {et~al.}(2021{\natexlab{a}})\citenamefont
  {Ni}, \citenamefont {Haglund}, \citenamefont {Wang}, \citenamefont {Xu},
  \citenamefont {Bernhard}, \citenamefont {Mandrus}, \citenamefont {Qian},
  \citenamefont {Mele}, \citenamefont {Kane},\ and\ \citenamefont
  {Wu}}]{ninatnano21}%
  \BibitemOpen
  \bibfield  {author} {\bibinfo {author} {\bibfnamefont {Zhuoliang}\
  \bibnamefont {Ni}}, \bibinfo {author} {\bibfnamefont {AV}~\bibnamefont
  {Haglund}}, \bibinfo {author} {\bibfnamefont {H}~\bibnamefont {Wang}},
  \bibinfo {author} {\bibfnamefont {B}~\bibnamefont {Xu}}, \bibinfo {author}
  {\bibfnamefont {C}~\bibnamefont {Bernhard}}, \bibinfo {author} {\bibfnamefont
  {DG}~\bibnamefont {Mandrus}}, \bibinfo {author} {\bibfnamefont
  {X}~\bibnamefont {Qian}}, \bibinfo {author} {\bibfnamefont {EJ}~\bibnamefont
  {Mele}}, \bibinfo {author} {\bibfnamefont {CL}~\bibnamefont {Kane}}, \ and\
  \bibinfo {author} {\bibfnamefont {Liang}\ \bibnamefont {Wu}},\ }\bibfield
  {title} {\enquote {\bibinfo {title} {{Imaging the N{\'e}el vector switching
  in the monolayer antiferromagnet {MnPSe$_3$} with strain-controlled Ising
  order}},}\ }\href {\doibase https://doi.org/10.1038/s41565-021-00885-5}
  {\bibfield  {journal} {\bibinfo  {journal} {Nat. Nanotechnol.}\ }\textbf
  {\bibinfo {volume} {16}},\ \bibinfo {pages} {782--787} (\bibinfo {year}
  {2021}{\natexlab{a}})}\BibitemShut {NoStop}%
\bibitem [{\citenamefont {Qian}\ \emph {et~al.}(submitted on Jan. 12,
  2022)\citenamefont {Qian}, \citenamefont {Occhialini}, \citenamefont {Emre},
  \citenamefont {Batyr}, \citenamefont {Kenji}, \citenamefont {Takashi},
  \citenamefont {Nuh},\ and\ \citenamefont {Riccardo}}]{songarxiv21}%
  \BibitemOpen
  \bibfield  {author} {\bibinfo {author} {\bibfnamefont {Song}\ \bibnamefont
  {Qian}}, \bibinfo {author} {\bibfnamefont {Connor~A.}\ \bibnamefont
  {Occhialini}}, \bibinfo {author} {\bibfnamefont {Ergecen}\ \bibnamefont
  {Emre}}, \bibinfo {author} {\bibfnamefont {Ilyas}\ \bibnamefont {Batyr}},
  \bibinfo {author} {\bibfnamefont {Watanabe}\ \bibnamefont {Kenji}}, \bibinfo
  {author} {\bibfnamefont {Taniguchi}\ \bibnamefont {Takashi}}, \bibinfo
  {author} {\bibfnamefont {Gedik}\ \bibnamefont {Nuh}}, \ and\ \bibinfo
  {author} {\bibfnamefont {Comin}\ \bibnamefont {Riccardo}},\ }\bibfield
  {title} {\enquote {\bibinfo {title} {{Experimental realization of a
  single-layer multiferroic}},}\ }\href {https://arxiv.org/pdf/2106.07661.pdf}
  {\bibfield  {journal} {\bibinfo  {journal} {arXiv}\ } (\bibinfo {year}
  {submitted on Jan. 12, 2022})},\ \Eprint {http://arxiv.org/abs/2106.07661}
  {2106.07661} \BibitemShut {NoStop}%
\bibitem [{\citenamefont {Ju}\ \emph {et~al.}(2021)\citenamefont {Ju},
  \citenamefont {Lee}, \citenamefont {Kim}, \citenamefont {Choi}, \citenamefont
  {Roh}, \citenamefont {Son}, \citenamefont {Park}, \citenamefont {Kim},
  \citenamefont {Jung}, \citenamefont {Kim} \emph {et~al.}}]{junanolett21}%
  \BibitemOpen
  \bibfield  {author} {\bibinfo {author} {\bibfnamefont {Hwiin}\ \bibnamefont
  {Ju}}, \bibinfo {author} {\bibfnamefont {Youjin}\ \bibnamefont {Lee}},
  \bibinfo {author} {\bibfnamefont {Kwang-Tak}\ \bibnamefont {Kim}}, \bibinfo
  {author} {\bibfnamefont {In~Hyeok}\ \bibnamefont {Choi}}, \bibinfo {author}
  {\bibfnamefont {Chang~Jae}\ \bibnamefont {Roh}}, \bibinfo {author}
  {\bibfnamefont {Suhan}\ \bibnamefont {Son}}, \bibinfo {author} {\bibfnamefont
  {Pyeongjae}\ \bibnamefont {Park}}, \bibinfo {author} {\bibfnamefont {Jae~Ha}\
  \bibnamefont {Kim}}, \bibinfo {author} {\bibfnamefont {Taek~Sun}\
  \bibnamefont {Jung}}, \bibinfo {author} {\bibfnamefont {Jae~Hoon}\
  \bibnamefont {Kim}},  \emph {et~al.},\ }\bibfield  {title} {\enquote
  {\bibinfo {title} {{Possible persistence of multiferroic order down to
  bilayer limit of van der Waals material NiI$_2$}},}\ }\href
  {doi.org/10.1021/acs.nanolett.1c01095} {\bibfield  {journal} {\bibinfo
  {journal} {Nano Lett.}\ }\textbf {\bibinfo {volume} {21}},\ \bibinfo {pages}
  {5126--5132} (\bibinfo {year} {2021})}\BibitemShut {NoStop}%
\bibitem [{\citenamefont {Bloembergen}\ and\ \citenamefont
  {Pershan}(1962)}]{bloembergenpr1962}%
  \BibitemOpen
  \bibfield  {author} {\bibinfo {author} {\bibfnamefont {N.}~\bibnamefont
  {Bloembergen}}\ and\ \bibinfo {author} {\bibfnamefont {P.~S.}\ \bibnamefont
  {Pershan}},\ }\bibfield  {title} {\enquote {\bibinfo {title} {Light waves at
  the boundary of nonlinear media},}\ }\href {\doibase 10.1103/PhysRev.128.606}
  {\bibfield  {journal} {\bibinfo  {journal} {Phys. Rev.}\ }\textbf {\bibinfo
  {volume} {128}},\ \bibinfo {pages} {606--622} (\bibinfo {year}
  {1962})}\BibitemShut {NoStop}%
\bibitem [{\citenamefont {Guyot-Sionnest}\ \emph {et~al.}(1986)\citenamefont
  {Guyot-Sionnest}, \citenamefont {Chen},\ and\ \citenamefont
  {Shen}}]{guyotprb1986}%
  \BibitemOpen
  \bibfield  {author} {\bibinfo {author} {\bibfnamefont {P.}~\bibnamefont
  {Guyot-Sionnest}}, \bibinfo {author} {\bibfnamefont {W.}~\bibnamefont
  {Chen}}, \ and\ \bibinfo {author} {\bibfnamefont {Y.~R.}\ \bibnamefont
  {Shen}},\ }\bibfield  {title} {\enquote {\bibinfo {title} {General
  considerations on optical second-harmonic generation from surfaces and
  interfaces},}\ }\href {\doibase 10.1103/PhysRevB.33.8254} {\bibfield
  {journal} {\bibinfo  {journal} {Phys. Rev. B}\ }\textbf {\bibinfo {volume}
  {33}},\ \bibinfo {pages} {8254--8263} (\bibinfo {year} {1986})}\BibitemShut
  {NoStop}%
\bibitem [{\citenamefont {Bloembergen}\ \emph {et~al.}(1966)\citenamefont
  {Bloembergen}, \citenamefont {Chang},\ and\ \citenamefont
  {Lee}}]{bloembergenprl1966}%
  \BibitemOpen
  \bibfield  {author} {\bibinfo {author} {\bibfnamefont {N.}~\bibnamefont
  {Bloembergen}}, \bibinfo {author} {\bibfnamefont {R.~K.}\ \bibnamefont
  {Chang}}, \ and\ \bibinfo {author} {\bibfnamefont {C.~H.}\ \bibnamefont
  {Lee}},\ }\bibfield  {title} {\enquote {\bibinfo {title} {Second-harmonic
  generation of light in reflection from media with inversion symmetry},}\
  }\href {\doibase 10.1103/PhysRevLett.16.986} {\bibfield  {journal} {\bibinfo
  {journal} {Phys. Rev. Lett.}\ }\textbf {\bibinfo {volume} {16}},\ \bibinfo
  {pages} {986--989} (\bibinfo {year} {1966})}\BibitemShut {NoStop}%
\bibitem [{\citenamefont {Lee}\ \emph {et~al.}(1967)\citenamefont {Lee},
  \citenamefont {Chang},\ and\ \citenamefont {Bloembergen}}]{leeprl1967}%
  \BibitemOpen
  \bibfield  {author} {\bibinfo {author} {\bibfnamefont {C.~H.}\ \bibnamefont
  {Lee}}, \bibinfo {author} {\bibfnamefont {R.~K.}\ \bibnamefont {Chang}}, \
  and\ \bibinfo {author} {\bibfnamefont {N.}~\bibnamefont {Bloembergen}},\
  }\bibfield  {title} {\enquote {\bibinfo {title} {Nonlinear electroreflectance
  in silicon and silver},}\ }\href {\doibase 10.1103/PhysRevLett.18.167}
  {\bibfield  {journal} {\bibinfo  {journal} {Phys. Rev. Lett.}\ }\textbf
  {\bibinfo {volume} {18}},\ \bibinfo {pages} {167--170} (\bibinfo {year}
  {1967})}\BibitemShut {NoStop}%
\bibitem [{\citenamefont {Reif}\ \emph {et~al.}(1991)\citenamefont {Reif},
  \citenamefont {Zink}, \citenamefont {Schneider},\ and\ \citenamefont
  {Kirschner}}]{reifprl1991}%
  \BibitemOpen
  \bibfield  {author} {\bibinfo {author} {\bibfnamefont {J.}~\bibnamefont
  {Reif}}, \bibinfo {author} {\bibfnamefont {J.~C.}\ \bibnamefont {Zink}},
  \bibinfo {author} {\bibfnamefont {C.-M.}\ \bibnamefont {Schneider}}, \ and\
  \bibinfo {author} {\bibfnamefont {J.}~\bibnamefont {Kirschner}},\ }\bibfield
  {title} {\enquote {\bibinfo {title} {Effects of surface magnetism on optical
  second harmonic generation},}\ }\href {\doibase 10.1103/PhysRevLett.67.2878}
  {\bibfield  {journal} {\bibinfo  {journal} {Phys. Rev. Lett.}\ }\textbf
  {\bibinfo {volume} {67}},\ \bibinfo {pages} {2878--2881} (\bibinfo {year}
  {1991})}\BibitemShut {NoStop}%
\bibitem [{\citenamefont {Reif}\ \emph {et~al.}(1993)\citenamefont {Reif},
  \citenamefont {Rau},\ and\ \citenamefont {Matthias}}]{reifprl1993}%
  \BibitemOpen
  \bibfield  {author} {\bibinfo {author} {\bibfnamefont {J.}~\bibnamefont
  {Reif}}, \bibinfo {author} {\bibfnamefont {C.}~\bibnamefont {Rau}}, \ and\
  \bibinfo {author} {\bibfnamefont {E.}~\bibnamefont {Matthias}},\ }\bibfield
  {title} {\enquote {\bibinfo {title} {Influence of magnetism on second
  harmonic generation},}\ }\href {\doibase 10.1103/PhysRevLett.71.1931}
  {\bibfield  {journal} {\bibinfo  {journal} {Phys. Rev. Lett.}\ }\textbf
  {\bibinfo {volume} {71}},\ \bibinfo {pages} {1931--1934} (\bibinfo {year}
  {1993})}\BibitemShut {NoStop}%
\bibitem [{\citenamefont {Chen}\ \emph {et~al.}(1981)\citenamefont {Chen},
  \citenamefont {Heinz}, \citenamefont {Ricard},\ and\ \citenamefont
  {Shen}}]{chenprl1981}%
  \BibitemOpen
  \bibfield  {author} {\bibinfo {author} {\bibfnamefont {C.~K.}\ \bibnamefont
  {Chen}}, \bibinfo {author} {\bibfnamefont {T.~F.}\ \bibnamefont {Heinz}},
  \bibinfo {author} {\bibfnamefont {D.}~\bibnamefont {Ricard}}, \ and\ \bibinfo
  {author} {\bibfnamefont {Y.~R.}\ \bibnamefont {Shen}},\ }\bibfield  {title}
  {\enquote {\bibinfo {title} {{Detection of molecular monolayers by optical
  second-harmonic generation}},}\ }\href {\doibase 10.1103/PhysRevLett.46.1010}
  {\bibfield  {journal} {\bibinfo  {journal} {Phys. Rev. Lett.}\ }\textbf
  {\bibinfo {volume} {46}},\ \bibinfo {pages} {1010--1012} (\bibinfo {year}
  {1981})}\BibitemShut {NoStop}%
\bibitem [{\citenamefont {Heinz}\ \emph {et~al.}(1982)\citenamefont {Heinz},
  \citenamefont {Chen}, \citenamefont {Ricard},\ and\ \citenamefont
  {Shen}}]{heinzprl82}%
  \BibitemOpen
  \bibfield  {author} {\bibinfo {author} {\bibfnamefont {T.~F.}\ \bibnamefont
  {Heinz}}, \bibinfo {author} {\bibfnamefont {C.~K.}\ \bibnamefont {Chen}},
  \bibinfo {author} {\bibfnamefont {D.}~\bibnamefont {Ricard}}, \ and\ \bibinfo
  {author} {\bibfnamefont {Y.~R.}\ \bibnamefont {Shen}},\ }\bibfield  {title}
  {\enquote {\bibinfo {title} {Spectroscopy of molecular monolayers by resonant
  second-harmonic generation},}\ }\href {\doibase 10.1103/PhysRevLett.48.478}
  {\bibfield  {journal} {\bibinfo  {journal} {Phys. Rev. Lett.}\ }\textbf
  {\bibinfo {volume} {48}},\ \bibinfo {pages} {478--481} (\bibinfo {year}
  {1982})}\BibitemShut {NoStop}%
\bibitem [{\citenamefont {Joy}\ and\ \citenamefont
  {Vasudevan}(1992)}]{joyprb1992}%
  \BibitemOpen
  \bibfield  {author} {\bibinfo {author} {\bibfnamefont {P.~A.}\ \bibnamefont
  {Joy}}\ and\ \bibinfo {author} {\bibfnamefont {S.}~\bibnamefont
  {Vasudevan}},\ }\bibfield  {title} {\enquote {\bibinfo {title} {{Magnetism in
  the layered transition-metal thiophosphates M${\mathrm{PS}}_{3}$ (M=Mn, Fe,
  and Ni)}},}\ }\href {\doibase 10.1103/PhysRevB.46.5425} {\bibfield  {journal}
  {\bibinfo  {journal} {Phys. Rev. B}\ }\textbf {\bibinfo {volume} {46}},\
  \bibinfo {pages} {5425--5433} (\bibinfo {year} {1992})}\BibitemShut {NoStop}%
\bibitem [{\citenamefont {Li}\ \emph {et~al.}(2013{\natexlab{b}})\citenamefont
  {Li}, \citenamefont {Cao}, \citenamefont {Niu}, \citenamefont {Shi},\ and\
  \citenamefont {Feng}}]{Lipnas13}%
  \BibitemOpen
  \bibfield  {author} {\bibinfo {author} {\bibfnamefont {Xiao}\ \bibnamefont
  {Li}}, \bibinfo {author} {\bibfnamefont {Ting}\ \bibnamefont {Cao}}, \bibinfo
  {author} {\bibfnamefont {Qian}\ \bibnamefont {Niu}}, \bibinfo {author}
  {\bibfnamefont {Junren}\ \bibnamefont {Shi}}, \ and\ \bibinfo {author}
  {\bibfnamefont {Ji}~\bibnamefont {Feng}},\ }\bibfield  {title} {\enquote
  {\bibinfo {title} {{Coupling the valley degree of freedom to
  antiferromagnetic order}},}\ }\href {\doibase 10.1073/pnas.1219420110}
  {\bibfield  {journal} {\bibinfo  {journal} {Proc. Natl. Acad. Sci. U. S. A.}\
  }\textbf {\bibinfo {volume} {110}},\ \bibinfo {pages} {3738--3742} (\bibinfo
  {year} {2013}{\natexlab{b}})}\BibitemShut {NoStop}%
\bibitem [{\citenamefont {Sivadas}\ \emph {et~al.}(2015)\citenamefont
  {Sivadas}, \citenamefont {Daniels}, \citenamefont {Swendsen}, \citenamefont
  {Okamoto},\ and\ \citenamefont {Xiao}}]{sivadasprb15}%
  \BibitemOpen
  \bibfield  {author} {\bibinfo {author} {\bibfnamefont {Nikhil}\ \bibnamefont
  {Sivadas}}, \bibinfo {author} {\bibfnamefont {Matthew~W.}\ \bibnamefont
  {Daniels}}, \bibinfo {author} {\bibfnamefont {Robert~H.}\ \bibnamefont
  {Swendsen}}, \bibinfo {author} {\bibfnamefont {Satoshi}\ \bibnamefont
  {Okamoto}}, \ and\ \bibinfo {author} {\bibfnamefont {Di}~\bibnamefont
  {Xiao}},\ }\bibfield  {title} {\enquote {\bibinfo {title} {Magnetic ground
  state of semiconducting transition-metal trichalcogenide monolayers},}\
  }\href {\doibase 10.1103/PhysRevB.91.235425} {\bibfield  {journal} {\bibinfo
  {journal} {Phys. Rev. B}\ }\textbf {\bibinfo {volume} {91}},\ \bibinfo
  {pages} {235425} (\bibinfo {year} {2015})}\BibitemShut {NoStop}%
\bibitem [{\citenamefont {Chittari}\ \emph {et~al.}(2016)\citenamefont
  {Chittari}, \citenamefont {Park}, \citenamefont {Lee}, \citenamefont {Han},
  \citenamefont {MacDonald}, \citenamefont {Hwang},\ and\ \citenamefont
  {Jung}}]{chittariprb16}%
  \BibitemOpen
  \bibfield  {author} {\bibinfo {author} {\bibfnamefont {Bheema~Lingam}\
  \bibnamefont {Chittari}}, \bibinfo {author} {\bibfnamefont {Youngju}\
  \bibnamefont {Park}}, \bibinfo {author} {\bibfnamefont {Dongkyu}\
  \bibnamefont {Lee}}, \bibinfo {author} {\bibfnamefont {Moonsup}\ \bibnamefont
  {Han}}, \bibinfo {author} {\bibfnamefont {Allan~H.}\ \bibnamefont
  {MacDonald}}, \bibinfo {author} {\bibfnamefont {Euyheon}\ \bibnamefont
  {Hwang}}, \ and\ \bibinfo {author} {\bibfnamefont {Jeil}\ \bibnamefont
  {Jung}},\ }\bibfield  {title} {\enquote {\bibinfo {title} {Electronic and
  magnetic properties of single-layer $m\mathrm{P}{X}_{3}$ metal phosphorous
  trichalcogenides},}\ }\href {\doibase 10.1103/PhysRevB.94.184428} {\bibfield
  {journal} {\bibinfo  {journal} {Phys. Rev. B}\ }\textbf {\bibinfo {volume}
  {94}},\ \bibinfo {pages} {184428} (\bibinfo {year} {2016})}\BibitemShut
  {NoStop}%
\bibitem [{\citenamefont {Kurosawa}\ \emph {et~al.}(1983)\citenamefont
  {Kurosawa}, \citenamefont {Saito},\ and\ \citenamefont
  {Yamaguchi}}]{kurosawajpsj1983}%
  \BibitemOpen
  \bibfield  {author} {\bibinfo {author} {\bibfnamefont {Ko}~\bibnamefont
  {Kurosawa}}, \bibinfo {author} {\bibfnamefont {Shozo}\ \bibnamefont {Saito}},
  \ and\ \bibinfo {author} {\bibfnamefont {Yasuo}\ \bibnamefont {Yamaguchi}},\
  }\bibfield  {title} {\enquote {\bibinfo {title} {{Neutron Diffraction Study
  on MnPS$_3$ and FePS$_3$}},}\ }\href {\doibase 10.1143/JPSJ.52.3919}
  {\bibfield  {journal} {\bibinfo  {journal} {Journal of the Physical Society
  of Japan}\ }\textbf {\bibinfo {volume} {52}},\ \bibinfo {pages} {3919--3926}
  (\bibinfo {year} {1983})}\BibitemShut {NoStop}%
\bibitem [{\citenamefont {Rule}\ \emph {et~al.}(2007)\citenamefont {Rule},
  \citenamefont {McIntyre}, \citenamefont {Kennedy},\ and\ \citenamefont
  {Hicks}}]{ruleprb07}%
  \BibitemOpen
  \bibfield  {author} {\bibinfo {author} {\bibfnamefont {K.~C.}\ \bibnamefont
  {Rule}}, \bibinfo {author} {\bibfnamefont {G.~J.}\ \bibnamefont {McIntyre}},
  \bibinfo {author} {\bibfnamefont {S.~J.}\ \bibnamefont {Kennedy}}, \ and\
  \bibinfo {author} {\bibfnamefont {T.~J.}\ \bibnamefont {Hicks}},\ }\bibfield
  {title} {\enquote {\bibinfo {title} {{Single-crystal and powder neutron
  diffraction experiments on $\mathrm{Fe}\mathrm{P}{\mathrm{S}}_{3}$: Search
  for the magnetic structure}},}\ }\href {\doibase 10.1103/PhysRevB.76.134402}
  {\bibfield  {journal} {\bibinfo  {journal} {Phys. Rev. B}\ }\textbf {\bibinfo
  {volume} {76}},\ \bibinfo {pages} {134402} (\bibinfo {year}
  {2007})}\BibitemShut {NoStop}%
\bibitem [{\citenamefont {Lancon}\ \emph {et~al.}(2016)\citenamefont {Lancon},
  \citenamefont {Walker}, \citenamefont {Ressouche}, \citenamefont {Ouladdiaf},
  \citenamefont {Rule}, \citenamefont {McIntyre}, \citenamefont {Hicks},
  \citenamefont {R\o{}nnow},\ and\ \citenamefont {Wildes}}]{lanconprb16}%
  \BibitemOpen
  \bibfield  {author} {\bibinfo {author} {\bibfnamefont {D.}~\bibnamefont
  {Lancon}}, \bibinfo {author} {\bibfnamefont {H.~C.}\ \bibnamefont {Walker}},
  \bibinfo {author} {\bibfnamefont {E.}~\bibnamefont {Ressouche}}, \bibinfo
  {author} {\bibfnamefont {B.}~\bibnamefont {Ouladdiaf}}, \bibinfo {author}
  {\bibfnamefont {K.~C.}\ \bibnamefont {Rule}}, \bibinfo {author}
  {\bibfnamefont {G.~J.}\ \bibnamefont {McIntyre}}, \bibinfo {author}
  {\bibfnamefont {T.~J.}\ \bibnamefont {Hicks}}, \bibinfo {author}
  {\bibfnamefont {H.~M.}\ \bibnamefont {R\o{}nnow}}, \ and\ \bibinfo {author}
  {\bibfnamefont {A.~R.}\ \bibnamefont {Wildes}},\ }\bibfield  {title}
  {\enquote {\bibinfo {title} {{Magnetic structure and magnon dynamics of the
  quasi-two-dimensional antiferromagnet ${\mathrm{FePS}}_{3}$}},}\ }\href
  {\doibase 10.1103/PhysRevB.94.214407} {\bibfield  {journal} {\bibinfo
  {journal} {Phys. Rev. B}\ }\textbf {\bibinfo {volume} {94}},\ \bibinfo
  {pages} {214407} (\bibinfo {year} {2016})}\BibitemShut {NoStop}%
\bibitem [{\citenamefont {Ressouche}\ \emph {et~al.}(2010)\citenamefont
  {Ressouche}, \citenamefont {Loire}, \citenamefont {Simonet}, \citenamefont
  {Ballou}, \citenamefont {Stunault},\ and\ \citenamefont
  {Wildes}}]{ressoucheprb10}%
  \BibitemOpen
  \bibfield  {author} {\bibinfo {author} {\bibfnamefont {E.}~\bibnamefont
  {Ressouche}}, \bibinfo {author} {\bibfnamefont {M.}~\bibnamefont {Loire}},
  \bibinfo {author} {\bibfnamefont {V.}~\bibnamefont {Simonet}}, \bibinfo
  {author} {\bibfnamefont {R.}~\bibnamefont {Ballou}}, \bibinfo {author}
  {\bibfnamefont {A.}~\bibnamefont {Stunault}}, \ and\ \bibinfo {author}
  {\bibfnamefont {A.}~\bibnamefont {Wildes}},\ }\bibfield  {title} {\enquote
  {\bibinfo {title} {{Magnetoelectric ${\text{MnPS}}_{3}$ as a candidate for
  ferrotoroidicity}},}\ }\href {\doibase 10.1103/PhysRevB.82.100408} {\bibfield
   {journal} {\bibinfo  {journal} {Phys. Rev. B}\ }\textbf {\bibinfo {volume}
  {82}},\ \bibinfo {pages} {100408} (\bibinfo {year} {2010})}\BibitemShut
  {NoStop}%
\bibitem [{\citenamefont {Ni}\ \emph {et~al.}(2021{\natexlab{b}})\citenamefont
  {Ni}, \citenamefont {Zhang}, \citenamefont {Hopper}, \citenamefont {Haglund},
  \citenamefont {Huang}, \citenamefont {Jariwala}, \citenamefont {Bassett},
  \citenamefont {Mandrus}, \citenamefont {Mele}, \citenamefont {Kane},\ and\
  \citenamefont {Wu}}]{niprl2021}%
  \BibitemOpen
  \bibfield  {author} {\bibinfo {author} {\bibfnamefont {Zhuoliang}\
  \bibnamefont {Ni}}, \bibinfo {author} {\bibfnamefont {Huiqin}\ \bibnamefont
  {Zhang}}, \bibinfo {author} {\bibfnamefont {David~A.}\ \bibnamefont
  {Hopper}}, \bibinfo {author} {\bibfnamefont {Amanda~V.}\ \bibnamefont
  {Haglund}}, \bibinfo {author} {\bibfnamefont {Nan}\ \bibnamefont {Huang}},
  \bibinfo {author} {\bibfnamefont {Deep}\ \bibnamefont {Jariwala}}, \bibinfo
  {author} {\bibfnamefont {Lee~C.}\ \bibnamefont {Bassett}}, \bibinfo {author}
  {\bibfnamefont {David~G.}\ \bibnamefont {Mandrus}}, \bibinfo {author}
  {\bibfnamefont {Eugene~J.}\ \bibnamefont {Mele}}, \bibinfo {author}
  {\bibfnamefont {Charles~L.}\ \bibnamefont {Kane}}, \ and\ \bibinfo {author}
  {\bibfnamefont {Liang}\ \bibnamefont {Wu}},\ }\bibfield  {title} {\enquote
  {\bibinfo {title} {Direct imaging of antiferromagnetic domains and anomalous
  layer-dependent mirror symmetry breaking in atomically thin
  ${\mathrm{mnps}}_{3}$},}\ }\href {\doibase 10.1103/PhysRevLett.127.187201}
  {\bibfield  {journal} {\bibinfo  {journal} {Phys. Rev. Lett.}\ }\textbf
  {\bibinfo {volume} {127}},\ \bibinfo {pages} {187201} (\bibinfo {year}
  {2021}{\natexlab{b}})}\BibitemShut {NoStop}%
\bibitem [{\citenamefont {Lee}\ \emph {et~al.}(2016)\citenamefont {Lee},
  \citenamefont {Lee}, \citenamefont {Ryoo}, \citenamefont {Kang},
  \citenamefont {Kim}, \citenamefont {Kim}, \citenamefont {Park}, \citenamefont
  {Park},\ and\ \citenamefont {Cheong}}]{leenanoletter16}%
  \BibitemOpen
  \bibfield  {author} {\bibinfo {author} {\bibfnamefont {Jae-Ung}\ \bibnamefont
  {Lee}}, \bibinfo {author} {\bibfnamefont {Sungmin}\ \bibnamefont {Lee}},
  \bibinfo {author} {\bibfnamefont {Ji~Hoon}\ \bibnamefont {Ryoo}}, \bibinfo
  {author} {\bibfnamefont {Soonmin}\ \bibnamefont {Kang}}, \bibinfo {author}
  {\bibfnamefont {Tae~Yun}\ \bibnamefont {Kim}}, \bibinfo {author}
  {\bibfnamefont {Pilkwang}\ \bibnamefont {Kim}}, \bibinfo {author}
  {\bibfnamefont {Cheol-Hwan}\ \bibnamefont {Park}}, \bibinfo {author}
  {\bibfnamefont {Je-Geun}\ \bibnamefont {Park}}, \ and\ \bibinfo {author}
  {\bibfnamefont {Hyeonsik}\ \bibnamefont {Cheong}},\ }\bibfield  {title}
  {\enquote {\bibinfo {title} {{Ising-type magnetic ordering in atomically thin
  FePS$_3$}},}\ }\href {\doibase 10.1021/acs.nanolett.6b03052} {\bibfield
  {journal} {\bibinfo  {journal} {Nano Lett.}\ }\textbf {\bibinfo {volume}
  {16}},\ \bibinfo {pages} {7433--7438} (\bibinfo {year} {2016})}\BibitemShut
  {NoStop}%
\bibitem [{\citenamefont {Wang}\ \emph {et~al.}(2016)\citenamefont {Wang},
  \citenamefont {Du}, \citenamefont {Liu}, \citenamefont {Hu}, \citenamefont
  {Zhang}, \citenamefont {Zhang}, \citenamefont {Owen}, \citenamefont {Lu},
  \citenamefont {Gan},\ and\ \citenamefont {Sengupta}}]{wang2dmat16}%
  \BibitemOpen
  \bibfield  {author} {\bibinfo {author} {\bibfnamefont {Xingzhi}\ \bibnamefont
  {Wang}}, \bibinfo {author} {\bibfnamefont {Kezhao}\ \bibnamefont {Du}},
  \bibinfo {author} {\bibfnamefont {Yu~Yang~Fredrik}\ \bibnamefont {Liu}},
  \bibinfo {author} {\bibfnamefont {Peng}\ \bibnamefont {Hu}}, \bibinfo
  {author} {\bibfnamefont {Jun}\ \bibnamefont {Zhang}}, \bibinfo {author}
  {\bibfnamefont {Qing}\ \bibnamefont {Zhang}}, \bibinfo {author}
  {\bibfnamefont {Man Hon~Samuel}\ \bibnamefont {Owen}}, \bibinfo {author}
  {\bibfnamefont {Xin}\ \bibnamefont {Lu}}, \bibinfo {author} {\bibfnamefont
  {Chee~Kwan}\ \bibnamefont {Gan}}, \ and\ \bibinfo {author} {\bibfnamefont
  {Pinaki}\ \bibnamefont {Sengupta}},\ }\bibfield  {title} {\enquote {\bibinfo
  {title} {{Raman spectroscopy of atomically thin two-dimensional magnetic iron
  phosphorus trisulfide (FePS$_3$) crystals}},}\ }\href
  {https://doi.org/10.1088/2053-1583/3/3/031009} {\bibfield  {journal}
  {\bibinfo  {journal} {2D Mater.}\ }\textbf {\bibinfo {volume} {3}},\ \bibinfo
  {pages} {031009} (\bibinfo {year} {2016})}\BibitemShut {NoStop}%
\bibitem [{\citenamefont {Zhang}\ \emph
  {et~al.}(2021{\natexlab{a}})\citenamefont {Zhang}, \citenamefont {Hwangbo},
  \citenamefont {Wang}, \citenamefont {Jiang}, \citenamefont {Chu},
  \citenamefont {Wen}, \citenamefont {Xiao},\ and\ \citenamefont
  {Xu}}]{zhangnanolett21}%
  \BibitemOpen
  \bibfield  {author} {\bibinfo {author} {\bibfnamefont {Qi}~\bibnamefont
  {Zhang}}, \bibinfo {author} {\bibfnamefont {Kyle}\ \bibnamefont {Hwangbo}},
  \bibinfo {author} {\bibfnamefont {Chong}\ \bibnamefont {Wang}}, \bibinfo
  {author} {\bibfnamefont {Qianni}\ \bibnamefont {Jiang}}, \bibinfo {author}
  {\bibfnamefont {Jiun-Haw}\ \bibnamefont {Chu}}, \bibinfo {author}
  {\bibfnamefont {Haidan}\ \bibnamefont {Wen}}, \bibinfo {author}
  {\bibfnamefont {Di}~\bibnamefont {Xiao}}, \ and\ \bibinfo {author}
  {\bibfnamefont {Xiaodong}\ \bibnamefont {Xu}},\ }\bibfield  {title} {\enquote
  {\bibinfo {title} {{Observation of giant optical linear dichroism in a zigzag
  antiferromagnet FePS$_3$}},}\ }\href {doi.org/10.1021/acs.nanolett.1c02188}
  {\bibfield  {journal} {\bibinfo  {journal} {Nano Lett.}\ }\textbf {\bibinfo
  {volume} {12}} (\bibinfo {year} {2021}{\natexlab{a}})}\BibitemShut {NoStop}%
\bibitem [{\citenamefont {Zhang}\ \emph
  {et~al.}(2021{\natexlab{b}})\citenamefont {Zhang}, \citenamefont {Jiang},
  \citenamefont {Lee}, \citenamefont {Lee}, \citenamefont {Mak},\ and\
  \citenamefont {Shan}}]{Xzhangnanolett21}%
  \BibitemOpen
  \bibfield  {author} {\bibinfo {author} {\bibfnamefont {Xiao-Xiao}\
  \bibnamefont {Zhang}}, \bibinfo {author} {\bibfnamefont {Shengwei}\
  \bibnamefont {Jiang}}, \bibinfo {author} {\bibfnamefont {Jinhwan}\
  \bibnamefont {Lee}}, \bibinfo {author} {\bibfnamefont {Changgu}\ \bibnamefont
  {Lee}}, \bibinfo {author} {\bibfnamefont {Kin~Fai}\ \bibnamefont {Mak}}, \
  and\ \bibinfo {author} {\bibfnamefont {Jie}\ \bibnamefont {Shan}},\
  }\bibfield  {title} {\enquote {\bibinfo {title} {{Spin dynamics slowdown near
  the antiferromagnetic critical point in atomically thin FePS$_3$}},}\
  }\href@noop {} {\bibfield  {journal} {\bibinfo  {journal} {Nano Lett.}\
  }\textbf {\bibinfo {volume} {21}},\ \bibinfo {pages} {5045--5052} (\bibinfo
  {year} {2021}{\natexlab{b}})}\BibitemShut {NoStop}%
\bibitem [{\citenamefont {Pan}\ \emph {et~al.}(1989)\citenamefont {Pan},
  \citenamefont {Wei},\ and\ \citenamefont {Shen}}]{panprb1989}%
  \BibitemOpen
  \bibfield  {author} {\bibinfo {author} {\bibfnamefont {Ru-Pin}\ \bibnamefont
  {Pan}}, \bibinfo {author} {\bibfnamefont {H.~D.}\ \bibnamefont {Wei}}, \ and\
  \bibinfo {author} {\bibfnamefont {Y.~R.}\ \bibnamefont {Shen}},\ }\bibfield
  {title} {\enquote {\bibinfo {title} {Optical second-harmonic generation from
  magnetized surfaces},}\ }\href {\doibase 10.1103/PhysRevB.39.1229} {\bibfield
   {journal} {\bibinfo  {journal} {Phys. Rev. B}\ }\textbf {\bibinfo {volume}
  {39}},\ \bibinfo {pages} {1229--1234} (\bibinfo {year} {1989})}\BibitemShut
  {NoStop}%
\bibitem [{\citenamefont {Zhang}\ \emph {et~al.}(submitted on Feb. 9,
  2022)\citenamefont {Zhang}, \citenamefont {Ni}, \citenamefont {Bai},
  \citenamefont {Peiris}, \citenamefont {Wu},\ and\ \citenamefont
  {Jariwala}}]{zhangarxiv2022}%
  \BibitemOpen
  \bibfield  {author} {\bibinfo {author} {\bibfnamefont {Huiqin}\ \bibnamefont
  {Zhang}}, \bibinfo {author} {\bibfnamefont {Zhuoliang}\ \bibnamefont {Ni}},
  \bibinfo {author} {\bibfnamefont {Aofeng}\ \bibnamefont {Bai}}, \bibinfo
  {author} {\bibfnamefont {Frank}\ \bibnamefont {Peiris}}, \bibinfo {author}
  {\bibfnamefont {Liang}\ \bibnamefont {Wu}}, \ and\ \bibinfo {author}
  {\bibfnamefont {Deep}\ \bibnamefont {Jariwala}},\ }\bibfield  {title}
  {\enquote {\bibinfo {title} {Cavity-enhanced linear dichroism in a van der
  waals antiferromagnet},}\ }\href {https://arxiv.org/abs/2202.04730} {\
  (\bibinfo {year} {submitted on Feb. 9, 2022})},\ \Eprint
  {http://arxiv.org/abs/2202.04730} {arXiv:2202.04730} \BibitemShut {NoStop}%
\end{thebibliography}
\end{document}


\title{Supplementary Material for\\
\normalsize{Observation of giant surface second harmonic generation coupled to nematic orders in the van der Waals antiferromagnet FePS$_3$}}

\author{Zhuoliang Ni}
\affiliation{Department of Physics and Astronomy, University of Pennsylvania, Philadelphia, Pennsylvania 19104, U.S.A}
\author{Nan Huang}
\affiliation{Department of Materials Science and Engineering, University of Tennessee, Knoxville, TN 37996, U.S.A.}
\author{Amanda V. Haglund}
\affiliation{Department of Materials Science and Engineering, University of Tennessee, Knoxville, TN 37996, U.S.A.}

\author{David G. Mandrus}
\affiliation{Department of Materials Science and Engineering, University of Tennessee, Knoxville, TN 37996, U.S.A.}
\affiliation{Materials Science and Technology Division, Oak Ridge National Laboratory, Oak Ridge, TN, 37831, U.S.A.}

\author{Liang Wu}
\email{liangwu@sas.upenn.edu}
\affiliation{Department of Physics and Astronomy, University of Pennsylvania, Philadelphia, Pennsylvania 19104, U.S.A}

\pacs{}
\maketitle


\section{Symmetry analysis of the surface SHG of FePS$_3$}
\subsection{$T>T_N$}

\begin{figure}[h]
\centering
\includegraphics[width=1\textwidth]{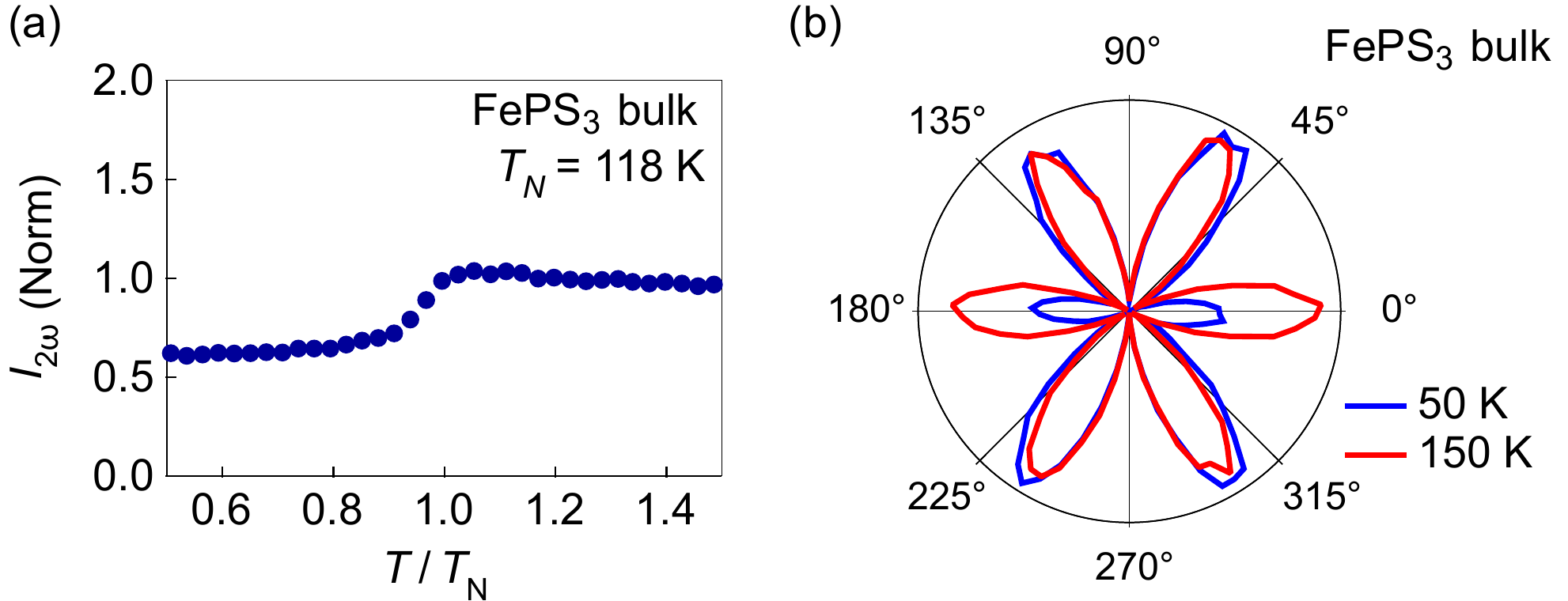}
\caption{ (a) Temperature dependence of SHG intensity (peak) in a bulk FePS$_3$ crystal (>10 $\mu$m). (b) Polarization-resolved SHG polar pattern measured at 50 K and 150 K. The fundamental light polarization and the detecting polarization are kept parallel when rotated simultaneously.}
\label{bulk}
\end{figure}

Above the N\'eel temperature, the point group of FePS$_3$ crystal is 2/m. In principle, both the surface term (without magnetization) and the quadruple term will contribute to the SHG. Since we do not see noticeable surface SHG in the parametric phase in our measurement, we neglect its contribution. Therefore, in the parametric phase, only the quadruple term is considered. Under normal incidence, we can simplify the EQ term as $\chi^{Q}_{ijzl}E_jk_zE_l (j,l=x,y)$. According to the mirror symmetry perpendicular to the b-axis we have $\chi^{EQ}_{xxzy}=\chi^{EQ}_{xyzx}=\chi^{EQ}_{yxzx}=\chi^{EQ}_{yyzy}=0$. In conclusion, the SHG intensity for the parallel configuration is
\begin{equation}
    I_{{\rm parallel}}(2\omega,T>T_N)\propto\left((\chi^{Q}_{xyzy}+2\chi^{Q}_{yxzy})\sin^2{\phi}\cos{\phi}+\chi^{Q}_{xxzx}\cos^3{\phi}\right)^2I^2(\omega),
\label{S1}
\end{equation}
where $I(\omega)$ is the incident power and $\phi$ is the angle between a-axis and the polarization of the incident light. As shown in Fig. \ref{bulk}(b), the polarization-resolved SHG polar pattern has a six-fold shape, which requires $\chi^{Q}_{xyzy}=\chi^{Q}_{yxzy}=\chi^{Q}_{xxzx}$ and thus a simpler form $I_{{\rm parallel}}(2\omega,T>T_N)\propto\left(\chi^{Q}_{xxzx}\cos{3\phi}\right)^2I^2(\omega)$.

Note the parallel SHG polar pattern in the paramagnetic phase peaks when incident light polarization is along the a-axis or 120$^\circ$ from the a-axis, which are three possible Zigzag directions. In the linear dichroism measurement on few-layer samples and bulk, we find when the reflection reaches its minimum, the parallel SHG pattern always reaches its peak, while the reflection reaches it maximum, the parallel SHG pattern always reaches its valley. From this observation, we conclude that the valley direction of the linear dichroism curve is the Zigzag direction. If the Zigzag direction is along the a-axis, then the direction with the lowest reflection is the a-axis, which is the 0$^{\circ}$ in Fig. \ref{bulk}(b). A recent comprehensive work of linear dichroism, X-ray diffraction and Raman scattering measurements combined with first principle calculation is consistent with our conclusion \cite{zhangnanolett21}. 

\subsection{$T<T_N$}

In the AFM phase, both magnetism-induced surface term and electric-quadruple term contribute to the SHG. Under normal incidence, the equation for surface SHG should be the same as Eq. \ref{S1} if the Zigzag direction is along the a-axis. If the Zigzag direction is 120$^\circ$ degrees from the a-axis, the system does not have the mirror symmetry any more, which means that the surface SHG patterns would also break mirror symmetry (see  Fig. 3. in the main text). 

\begin{figure}[htb]
\centering
\includegraphics[width=0.8\textwidth]{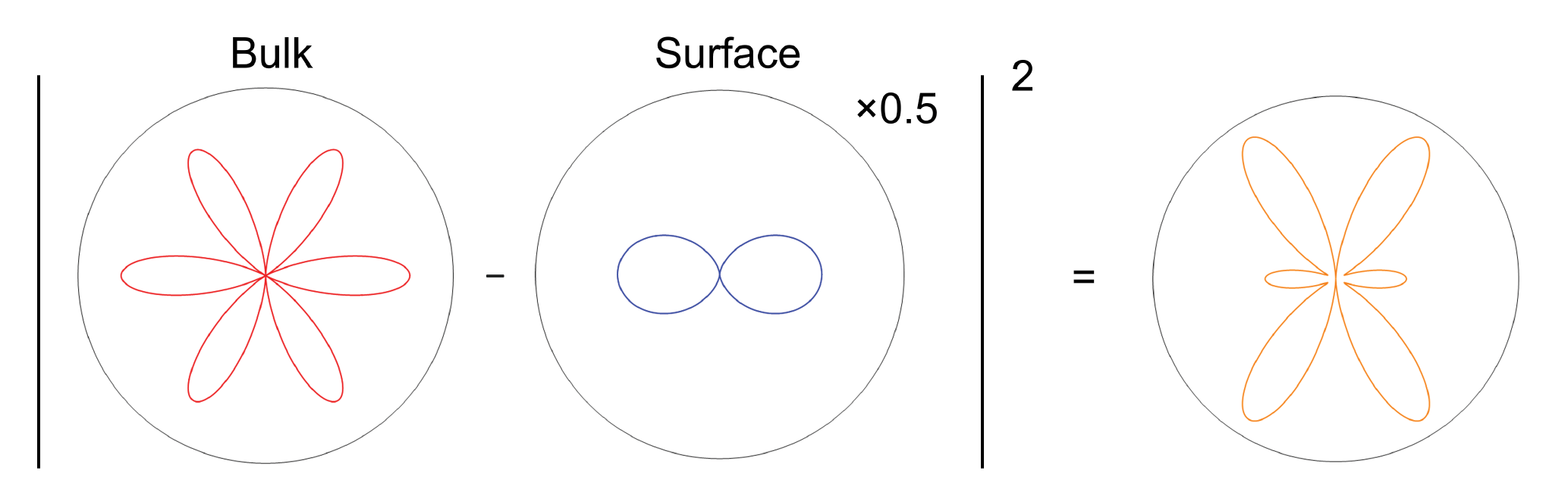}
\caption{Simulation of the interference between the bulk SHG and surface SHG.}
\label{bulksim}
\end{figure}

In our experiment, we see a dominant two-fold surface SHG polar pattern in few-layer FePS$_3$ samples and the node direction of the SHG is perpendicular to the Zigzag direction. To qualitatively explain the observed behaviour, we assume the surface has three-fold rotation as well as the mirror symmetry ($C_{3v}$). When Zigzag order forms, the three-fold rotation symmetry is broken, and the only left mirror symmetry is in the plane of the Zigzag direction and z-axis ($C_{1v}$). The symmetry allows limited elements in the surface SHG susceptibility tensor: $\chi_{yxy}^s=\chi_{yyx}^s$, $\chi_{xyy}^s$ and $\chi_{xxx}^s$. The polarization-dependent second-harmonic electric field in the parallel geometry is then given by

\begin{equation}
    E_{{\rm parallel}}^s(2\omega)\propto\left((\chi^{s}_{xyy}+2\chi^{s}_{yxy})\sin^2{\phi}\cos{\phi}+\chi^{s}_{xxx}\cos^3{\phi}\right)E^\red{2}(\omega),
\label{S2}
\end{equation}
where $\phi$ is the angle between the Zigzag direction and the polarization of the incident light. Note that no matter what the nonzero tensor elements are, there is always a node in the SHG polar pattern, which is perpendicular to the Zigzag direction. In the experiment, we observe a two-fold SHG pattern for magnetism-induced surface response, which implies that the tensor elements approximately satisfies $\chi^{s}_{xyy}=-\chi^{s}_{yxy}=-\chi^{s}_{xxx}$. If we further neglect the electric-quadruple term in few-layer samples, the polarization-dependent SHG intensity in parallel geometry would be $I_{{\rm parallel}}(2\omega,T<T_N)\propto (\chi^{s}_{xxx}\cos{\phi})^2I^2(\omega)$. In bulk samples, the SHG signal is the interference of surface term and electric-quadruple term (see Fig. \ref{bulk} (b) and a simulation in Fig.\ref{bulksim}).

\section{Estimation of the surface nonlinear susceptibility $\chi_{ijk}^s$ of the few-layer FePS$_3$}

We use GaAs (111) crystal under the same experiment condition as a reference for estimating second-order susceptibility. The peak of the six-fold SHG patterns of GaAs we get is around 18000 c.p.s. A much larger response is expected because more atoms are activated to produce SHG due to the thickness, and it has been known with large second-order susceptibility. To get the second-order susceptibility of FePS$_3$, we use the formula derived by Bloembergen and Pershan\cite{bloembergenpr1962} for the GaAs,
\begin{equation}
    \chi^{(2)}_R\equiv-\frac{E_R(2\omega)}{\epsilon_0E(\omega)^2}=\frac{\lambda}{2\pi}\frac{\chi^{(2)}}{(\epsilon^{1/2}(2\omega)+\epsilon^{1/2}(\omega))(\epsilon^{1/2}(2\omega)+1)}\left(\frac{2}{\epsilon^{1/2}(\omega)+1}\right)^2,
\end{equation}
where $\epsilon$ is the relative dielectric constant and $\lambda$ is the wavelength of the fundamental beam. For few-layer FePS$_3$ sample, a much simpler equation is applied:
\begin{equation}
    \chi_R^{(2)}\equiv-\frac{E_R(2\omega)}{\epsilon_0E(\omega)^2}=\frac{2\chi^{(2),s}}{\epsilon^{1/2}(2\omega)+1}\left(\frac{2}{\epsilon^{1/2}(\omega)+1}\right)^2.
\end{equation} 

Note $\chi^{(2),s}$ for surface SHG and $\chi^{(2)}$ for bulk electric-dipole SHG have different fundamental units due to their different definitions. With the refractive index of GaAs\cite{jellisonopticalmater1992} and FePS$_3$, and the second-harmonic susceptibility of GaAs\cite{bergfeldprl03} at room temperature, we estimate the magnitude of the surface SHG susceptibility of 2L-5L FePS$_3$ to be 0.08--0.13 nm$^2$/V.

\section{Zigzag domains measured in a  FePS$_3$ thick flake}
\begin{figure}[htb]
\centering
\includegraphics[width=0.6\textwidth]{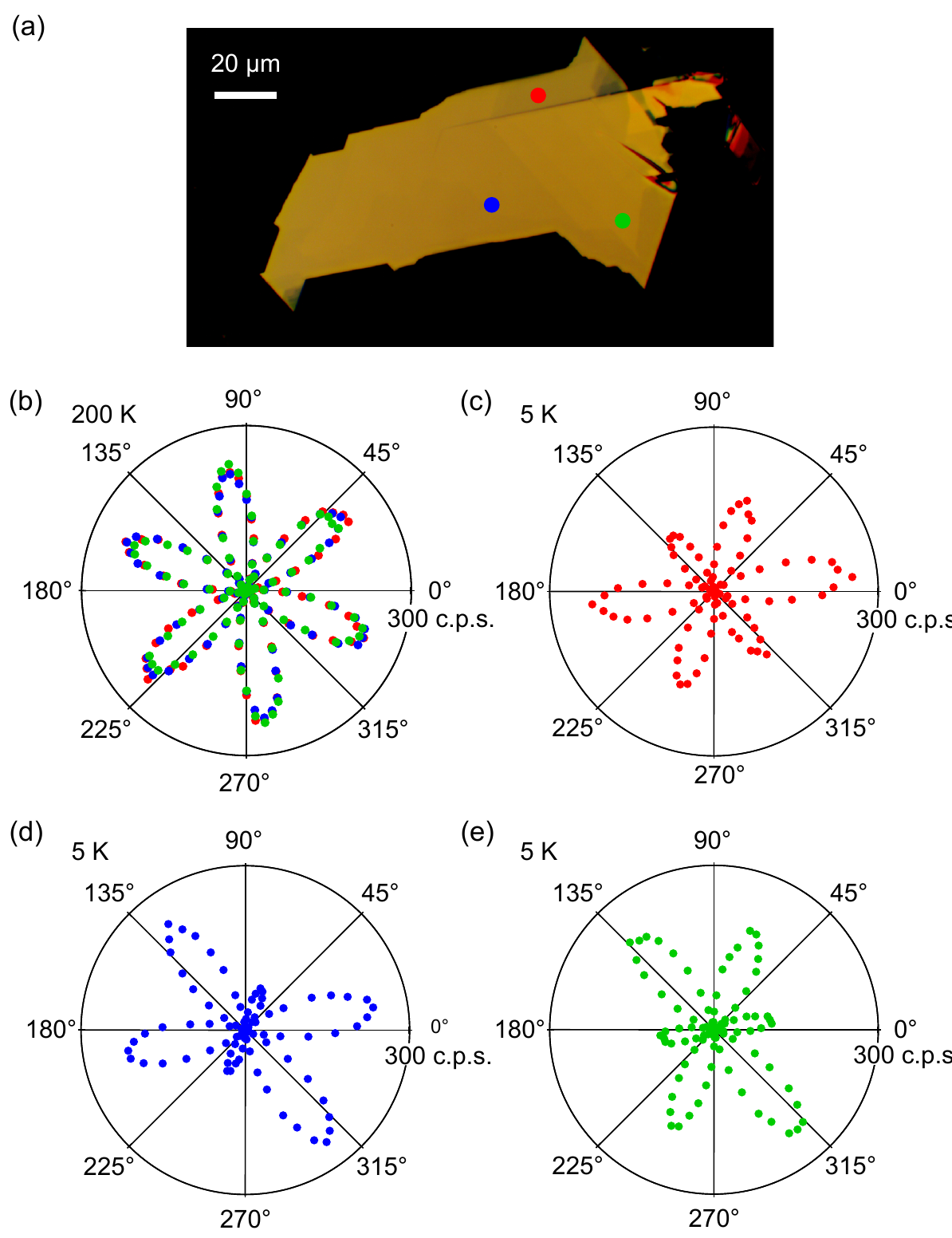}
\caption{ (a) Optical image of the FePS$_3$ thick flake. (b) Polarization-resolved SHG polar patterns measured at 200 K. The fundamental light polarization and the detecting polarization are kept parallel when rotated simultaneously. SHG signals from three different positions (labeled in (a) by color) on the flake are measured. (c-e) Polarization-resolved SHG polar patterns measured at 5 K on the same three points.}
\label{thickflake}
\end{figure}

We performed polarization-resolved SHG measurement on a $\approx$ 80 nm thick FePS$_3$ flake exfoliated on the SiO$_2$/Si wafer, as shown in Fig. \ref{thickflake}. Three different points are chosen. At 200 K, all the three points show the same six-fold SHG patterns (Fig.\ref{thickflake}(b)), and have the same LD data. At 5 K, however, the three points show different SHG responses. Reduction on lobes at different angles happens at different points, as shown in Fig. \ref{thickflake}(c--e). According to the previous section, the direction of the reduced lobes in the parallel SHG patterns indicates the direction of the Zigzag orders. Therefore, three different Zigzag directions coexist in this thick flake. Note the three points are from the different areas with different thickness, and thick samples tend to have stacking faults. Therefore, in order to investigate the origin of the three domains, we use a 3L sample without structure domains in the main text.


\section{Enhancement of the magnetism-coupled SHG in FePS$_3$ thin samples}
\begin{figure}
    \centering
    \includegraphics[width=0.9\textwidth]{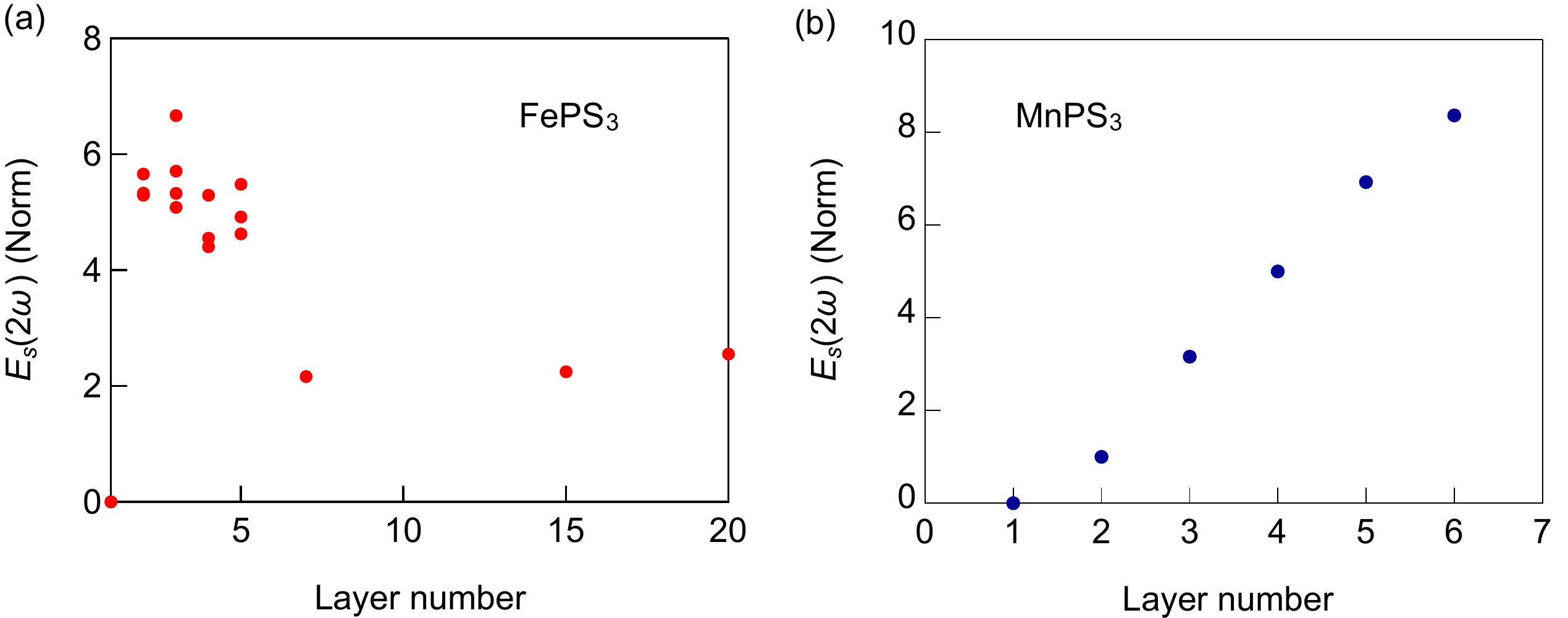}
    \caption{(a) Layer dependence of the extracted magetism-coupled second harmonic electric field in FePS$_3$. (b) Layer dependence of the magnetism-coupled second harmonic electric field in MnPS$_3$ from reference \cite{niprl2021}.}
    \label{layersum}
\end{figure}

From the previous discussion, we know the magnetism-coupled surface SHG is interfered with nonmagnetic SHG terms. Here we use the SHG signal measured at high temperature and low temperature to extract the magnetism-coupled second-harmonic electric field. The layer dependence of the second-harmonic electric field is shown in Figure \ref{layersum}(a). Though getting pure magnetism-coupled signal in this way might not be exactly accurate considering the related phase difference of two contribution is unknown, it is apparently there is an enhancement of the surface SHG in FePS$_3$ sample below 6 L. Below 6 L, the response is nearly consistent, suggesting its origin is from the surface. In contrast, Figure \ref{layersum}(b) shows a linear dependence of second-harmonic electric fields and the layer numbers in MnPS$_3$, where the magnetism-coupled signal is from bulk\cite{niprl2021}.

\section{More data on linear dichroism mapping of the 3 L FePS$_3$ sample}

\begin{figure}[htb]
\centering
\includegraphics[width=0.8\textwidth]{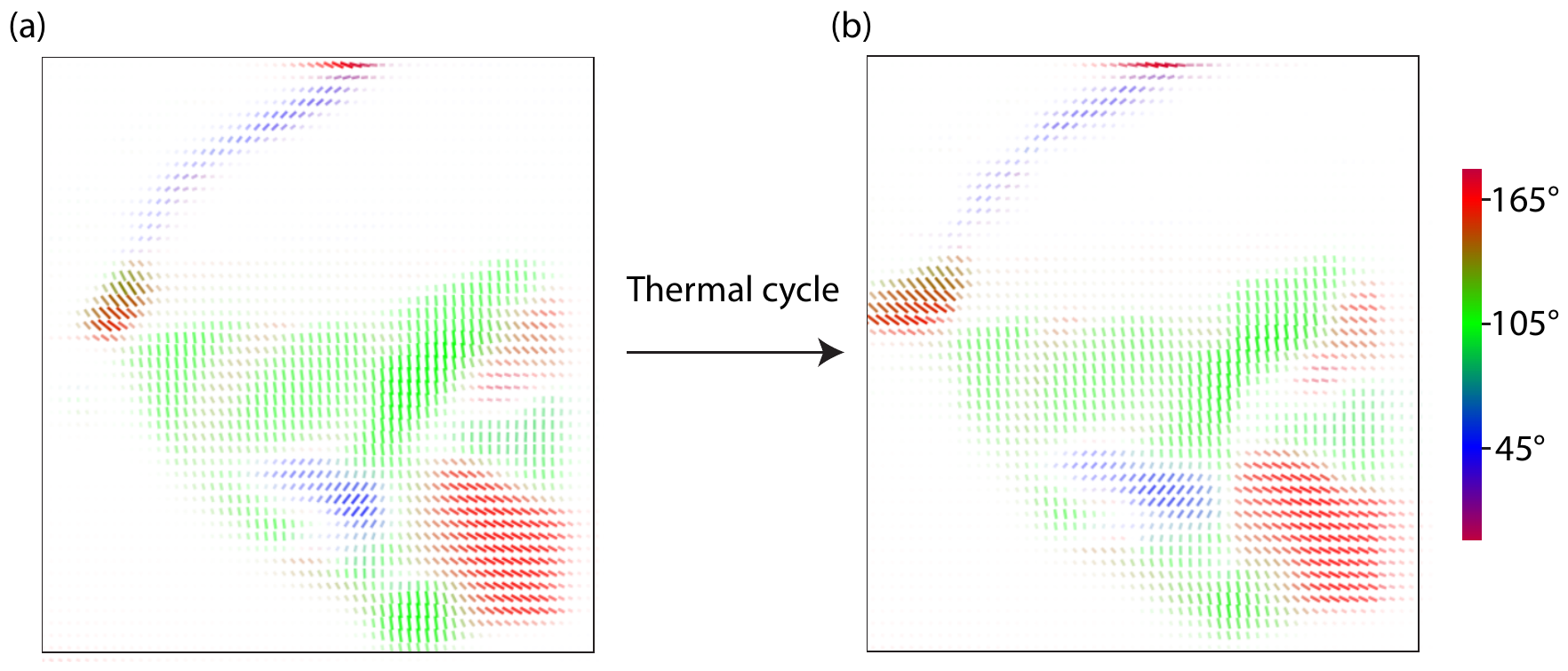}
\caption{ (a) Linear dichroism mapping of the multi-layer FePS$_3$ sample shown in main text Fig.3. The direction and the color of the segment at each point represent the valley positions of the linear dichroism curve. The measurement is done at 30 K.  (b) Linear dichroism mapping on the same sample at 30 K after a thermal cycle.}
\label{thermalcycle}
\end{figure}

We have shown that the Zigzag direction is not uniform across the 3 L sample at 30 K in the main text. To investigate whether direction distribution is random or fixed when the AFM order forms, we perform a thermal cycle and do a linear dichroism mapping again. Fig. \ref{thermalcycle} (a) and (b) represent the Zigzag direction distribution before and after the thermal cycle. No change of the Zigzag orientation is observed, which means the Zigzag direction distribution is fixed, probably by some external factors such as the strain from the substrate.

\section{Linear and nonlinear responses of the monolayer FePS$_3$}

\begin{figure}[htb]
\centering
\includegraphics[width=1\textwidth]{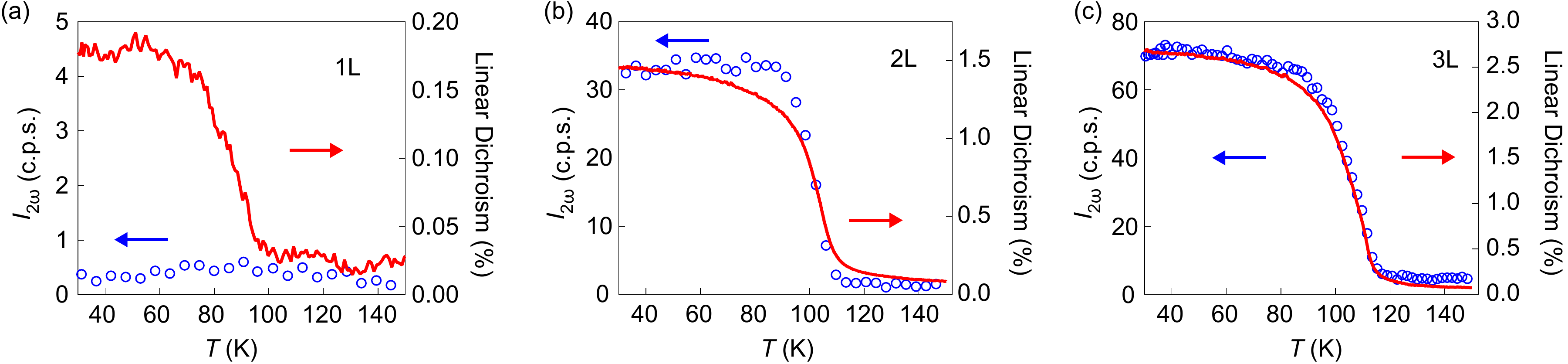}
\caption{ (a-c) Temperature dependence of SHG intensity and magnitude of the linear dichroism of (a) a monolayer, (b) a bilayer, and (c) a trilayer FePS$_3$ sample, respectively.}
\label{1L}
\end{figure}

In the main text we do not include the data from FePS$_3$ monolayer, mainly because we do not see any surface SHG signal in all the monolayer FePS$_3$. To minimize any possible air degradation, we prepare the samples in a glovebox (H$_2$O and O$_2$ < 0.5 ppm) and then quickly transfer them into a cryostat chamber under vacuum. The measured linear dichroism and SHG data are shown in Fig. \ref{1L}. Compared to bilayer and trilayer, the linear dichroism data on the monolayer is one order of magnitude smaller. We also see some decrease of the transition temperature. The strong suppression of the linear dichroism suggests the charge-spin coupling in the monolayer is reduced. In terms of the SHG measurement, we do not see any measurable change of the SHG signal across the N\'eel temperature. It is reasonable considering the linear dichroism becomes one order of magnitude smaller. If both the linear optical response $\chi^{(1)}$ and nonlinear response $\chi^{(2)}$ reduce one order of magnitude, the linear response (such as linear dichroism) would reduce one order while the second-order optical response (such as SHG intensity) would reduce two orders of magnitude, which requires measurement sensitivity below our detection limits \cite{ninatnano21,niprl2021}. Here, we show that the anti-ferromagnetic order persists down to the monolayer according to the LD measurement, which agrees with recent Raman scattering measurements\cite{leenanoletter16,wang2dmat16} and another LD work\cite{zhangnanolett21}. 

